\documentclass[11pt, a4paper]{article}
\usepackage[margin=25mm]{geometry}
\usepackage{cite}
\usepackage{graphicx}
\usepackage{amsmath,amsfonts}
\usepackage{algorithmic}
\usepackage{algorithm}
\usepackage{array}
\usepackage{stfloats}
\usepackage{url}
\usepackage{subfigure}
\providecommand{\keywords}[1]
{
  \small	
  \textbf{\textit{Keywords---}} #1
}

\begin{document}

\title{Terahertz imaging super-resolution for documental heritage diagnostics}

\author{Danae~Antunez~Vazquez, Laura~Pilozzi,  \\     Eugenio~DelRe, Claudio~Conti, and~Mauro~Missori%
        %<-this % stops a space
\thanks{This work was supported by Ministero dell’Università e della Ricerca (CN0000013 HPC, ARS01-00734). The work of Danae Antunez Vazquez was supported by the ARCHMAT European program. (Corresponding author: Mauro Missori, e-mail: mauro.missori@cnr.it). Danae Antunez Vazquez, Laura Pilozzi, and Mauro Missori are with the Institute for Complex Systems, National Research Council (ISC-CNR), Via dei Taurini 19, 00185 Rome, Italy;   Eugenio DelRe, Claudio Conti are with the Department of Physics, University Sapienza, Piazzale Aldo Moro 5, 00185 Rome, Italy.}% <-this % stops a space 
\thanks{Manuscript received April 9, 2024; revised ...}}

“This work has been submitted to the IEEE for possible publication. Copyright may be transferred without notice, after which this version may no longer be accessible.”

\maketitle

\begin{abstract}
Terahertz imaging provides valuable insights into the composition and structure of objects or materials, with applications spanning security screening, medical imaging, materials science, and cultural heritage preservation.
Despite its widespread utility, traditional terahertz imaging is limited in spatial resolution to approximately 1 mm according to Abbe's formula. In this paper, we propose a novel super-resolution method for terahertz time-domain spectroscopy systems. Our approach involves spatial filtering through scattering in the far-field of high spatial frequency components of the imaged sample. This method leverages evanescent wave filtering using a knife edge, akin to a standard structured illumination scheme.
We demonstrate improved spatial resolution in slit diffraction, edge imaging, and reflection imaging of structures fabricated on a paper substrate using commonly encountered materials in works of art and documents. Furthermore, we present super-resolved images of an ancient document on parchment, showcasing the effectiveness of our proposed method.

\end{abstract}


\section{Introduction}
Spatial resolution refers to the ability of an optical instrument to distinguish details of a physical object reproduced in an image. However, due to the wave nature of light, resolution is limited by diffraction, which imposes constraints on overcoming Abbe's limit \cite{Abbe}. This physical limitation, known as the diffraction limit, defines the minimum distance between two features that can be resolved with a certain contrast by optical instruments, with the specific value depending on the imaging scheme employed.

According to Abbe's theory of image formation, the lateral resolution (i.e., resolution in the image plane) for an ordinary microscope is given by $d_{inch}= 0.61\lambda/\text{NA}$ when using incoherent illumination, while a confocal scheme with coherent illumination \cite{Juang_1988} provides a diffraction-limited lateral resolution of $d_{ch}= 0.82\lambda/(\sqrt{2}\text{NA})$. Here, $\lambda$ represents the radiation wavelength and $\text{NA}$ denotes the numerical aperture of the objective. The numerical factors are determined by numerically evaluating the Airy function under the paraxial approximation, applying the half-width at half-maximum criterion \cite{Juang_1988}.

As a consequence of the diffraction limit, an acquired image is a blurred representation of the actual object under investigation. This blurring is described by the so-called Point Spread Function (PSF), which represents the response of a camera system to a point source, a ``spatial'' impulse.
Indeed, for every point in the object, the minimal size of its focal spot is finite, and the lateral and axial extents of the intensity distribution in the focal region are determined by the three-dimensional diffraction pattern of light emitted and transmitted to the image plane through the objective.
Due to the linearity of the image formation process, the acquired image, denoted as $I(x,y)$, is then a focused image of the sample $s(x,y)$ convoluted with the PSF $h(x,y)$, expressed as

\begin{equation}
	I(x,y)=s(x,y) \otimes h(x,y)
	\label{eq_convol1}
\end{equation}

As known, THz time-domain spectroscopy (THz-TDS) and THz pulsed imaging offer valuable insights into the composition and structure of various objects or materials, and their applications range from security screening and medical imaging to materials science and cultural heritage preservation \cite{book, Yang_2016, Wang_2017, Wang_2021, Koch_2023}.
Nevertheless, their spatial resolution often falls short of a detailed analysis since the wavelengths associated with THz radiation (100 - 1000 $\mu$m) impose stringent limitations. 
Relevant in image processing are techniques such as super-resolution, which can be employed to mitigate the effects of a broad PSF, particularly in the lower THz range where imaging quality may be significantly degraded by diffraction for details smaller than approximately one millimetre in size. 

Several methods have been proposed to improve the resolution and quality of THz images, each of them offering unique advantages and drawbacks \cite{Totero_Gongora_2020}. Among the methods for obtaining super-resolved images in the far-field, convolutional neural networks are a valuable super-resolution method due to their ability to capture complex patterns and features from input data \cite{Yuan, Lei} but require extensive training sets, a problem recently addressed in \cite{Ljubenovi_2023}. An alternative method is based on a confocal waveguide \cite{Yu}, an imaging technique providing improved resolution but involving complexity and influenced by sample properties.

A powerful and versatile super-resolution technique is Structured-Illumination Microscopy (SIM) \cite{SIM,Strohl:16,Gusta}, allowing for the retrieval of subwavelength features of an object by conversion of evanescent waves into propagating waves.
Given that the whole process of blurring is in fact a low-pass filtering, SIM encodes sample structural details, corresponding to high spatial frequencies, in low-frequency signals via spatial frequency mixing.

Its conventional implementation uses a pattern excitation $e(x)$, formed by the interference of laser beams on the sample so that $I(x,y)=[s(x,y) \times e(x,y)] \otimes h(x,y)$. The accessible frequency content results in a shifting of the sample information to new frequency positions centered at the spatial frequency of the excitation pattern. SIM then delivers better resolution than conventional microscopy by shifting high spatial frequencies into the accessible passband of the objective optical transfer function, the PSF Fourier transform, and can thus be viewed as a method that increases the optical transfer function support. 

In this work, we apply a further approach to THz imaging, the knife-edge (KE) technique \cite{Firester1977, sub_2015}, a super-resolution scheme that operates with a structured illumination plane at a subwavelength distance from the sample, to reduce the effects of the PSF and improve the spatial resolution of images of samples. 
Hence, we focus on a far-field super-resolution THz imaging system based on a freestanding KE and transmission or reflective confocal configuration for the THz beamline, as employed in THz-TDS systems \cite{metamaterials}.
We show enhanced spatial resolution in slit diffraction, edge imaging, and reflection imaging of structures created on paper using materials commonly found in artwork and documents. Additionally, we exhibit super-resolved images of an ancient parchment document, highlighting the efficacy of our proposed method.

\section{Methods}

\subsection{Knife-edge (KE) scan method}
\label{par:KE-method}
The KE scan method is a super-resolution scheme that operates with a structured illumination plane, practically obtained by a scanning blade at a variable sub-wavelength distance from the surface of the sample. It achieves super-resolution in a single-wavelength beam-profiling execution, unlike advanced super-resolution techniques that typically involve delivering and collecting radiation at different wavelengths, deconvolution calculations, fluorescence phenomena, and/or nonlinear interactions with other light fields \cite{confocal}. In the THz range, KE scans have been used to significantly enhance the spatial resolution of images of laser-induced broadband source points. Interestingly, an optically induced virtual KE technique was demonstrated using structured illumination with a visible laser in the terahertz-emitting (object) plane \cite{sub_2015}.

Assuming linearity in the reconstruction operation, the image of any object can be calculated by breaking the object into smaller components, imaging each of these components individually, and subsequently summing the outcomes. As the object is divided into increasingly smaller sections, it essentially consists of an array of infinitesimal point-like objects. Each of these point objects generates a PSF in the image, adjusted and scaled according to the corresponding point's location and intensity, respectively. Consequently, the resulting image represents an assembly of often overlapping PSFs. This image formation process is formally represented by the convolution equation (Eq. \ref{eq_convol1}): the object convolved with the PSF of the imaging setup yields the obtained image.

To simplify the analysis, we will consider the mechanism of super-resolution for a one-dimensional movement of the blade, with a blade in the plane $z = 0$, covering the optical field on $x' > 0$. Let's assume also that the THz radiation is monochromatic. The object plane and the sample surface are at $z = 0$, and the emitted or reflected radiation with sub-wavelength spatial features is in the direction $z > 0$. 

Consider the THz electric field reflected by the sample to be $E(x,z)$ (see inset of Fig.~\ref{RBL} for axis definitions), which after the insertion of the blade become $E'(x,z,x')=E(x,z)\theta(x-x')$, where $\theta(x-x')$ is the step function that represents an ideally sharp and opaque blade with its edge in $x'$ (edge diffraction effects are neglected).

Without the blade, the image collected in the far-field is formed by the convolution of the reflected field with the PSF (Eq. \ref{eq_convol1}), i.e.

\begin{equation}
	\operatorname{I}(x,z) = \int_{\mathbb{R}} E\left(u, z\right) \operatorname{PSF}(x-u) du
	\label{eq_convol2}
\end{equation}
\\
which is the convolution of the THz electric field distribution, $E(u,z)$ reflected by the sample with the PSF of the optical set-up.
With the blade, the image collected in the far-field is
\begin{equation}
	\operatorname{I_{SR}}(x,z,x')= \int_{\mathbb{R}} E\left(u, z, x'\right) \operatorname{PSF}(x-u) du
	\label{eq_image_blade}
\end{equation}
\\ 
which is the convolution of the THz electric field distribution, $E(u,z,x')$, immediately after the blade, with the PSF of the optical set-up.

The collection of Eq. \ref{eq_image_blade} for several positions of the blade edge $x'$ gives the integral of the transmitted field  $\int_{\mathbb{R}} dx \int_{\mathbb{R}} E\left(u, z, x'\right) \operatorname{PSF}(x-u) d u $. In order to obtain the scattered electric field distribution $E_{S}$ a differentiation is applied, providing: 
\begin{equation}
	E_{S}\left(x', z\right) =-\frac{d}{d x'} \int_{\mathbb{R}} d x \int_{\mathbb{R}} E\left(u, z, x'\right) \operatorname{PSF}(x-u) d u
\end{equation}
\\
Making use of the normalization of the PSF, we obtain:
\begin{equation}
	E_{S}\left(x', z\right) =-\int_{\mathbb{R}} E(u, z) \frac{d}{d x'} \theta\left(u-x'\right) d u=E\left(x', z\right)
\end{equation}
\\
so that the high spatial frequencies of the spectrum of $E(x',z)$ will be fully transferred to the reconstructed far-field image $E_S(x', z)$ which is, therefore, no longer limited by diffraction and is a remotely super-resolved image.

The scattering of the evanescent-wave intensity by the blade edge into propagating waves newly formed at $z > 0$ allows super-resolved image reconstruction in the far-field. 
This is achieved by subtracting, for each pixel in the image, the total far-field electric field collected at each blade position $x'$ from the field collected at the previous position $x' - dx'$, and taking the maximum of this difference as the image pixel intensity value.

\subsection{THz-TDS super-resolution imaging set-up} 
THz imaging was carried out employing a Menlo Systems (Germany) TERA K15 THz time-domain (THz-TDS) system equipped with photo-conductive dipole antenna (PCA) \cite{Burford_2017} exited by a femtosecond fibre-coupled laser (Menlo Systems T-Light) with 1560~nm emission wavelength. The PCA emits pulsed broadband THz radiation whose peak is at 0.3~THz.

\begin{figure}[h!]
	\centering
	\includegraphics[height=6.3cm]{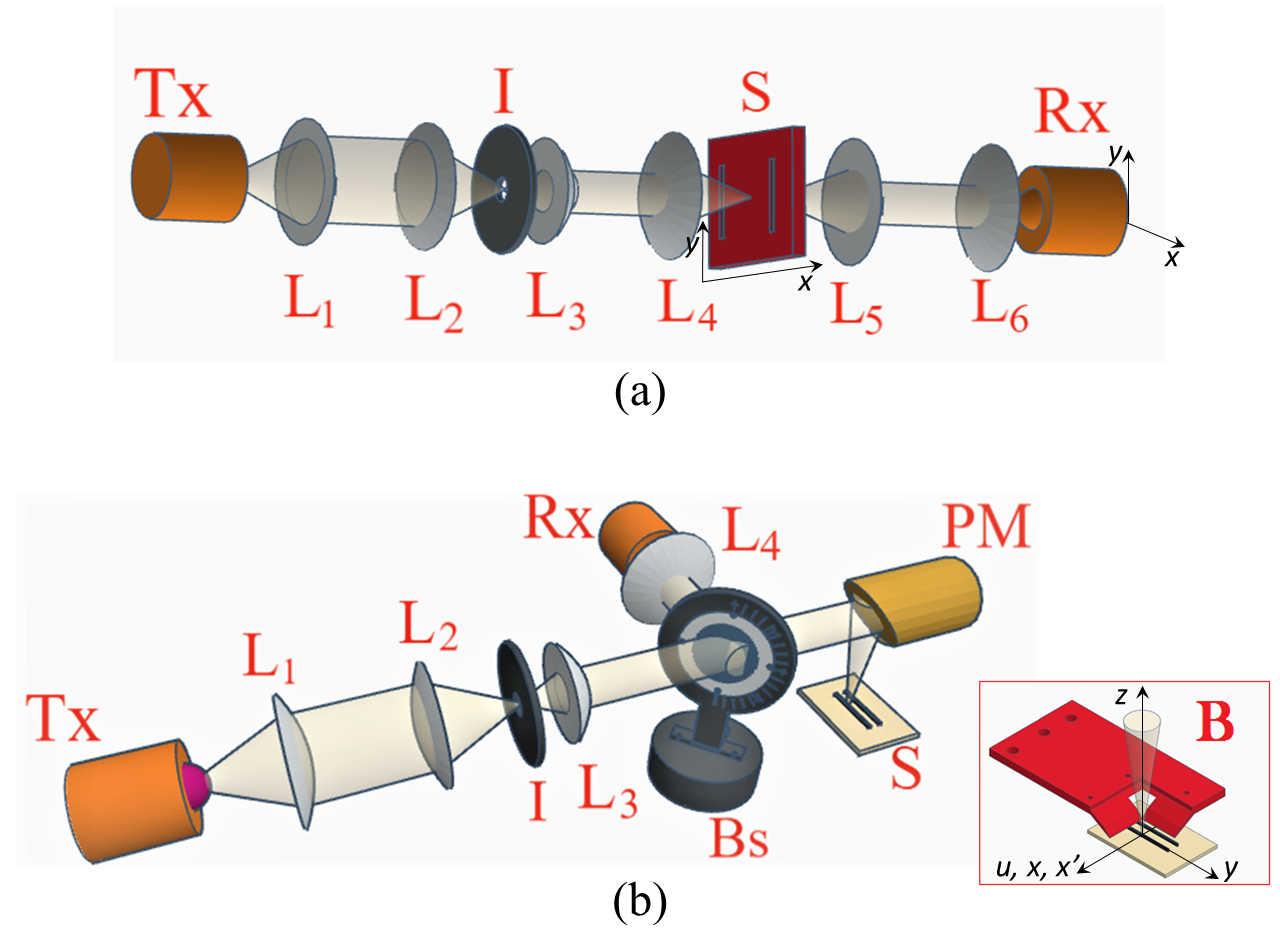}
	\caption{Transmission (a) and reflection (b) confocal configuration for the THz-TDS set-up. The blades (B) shown in the inset are placed after the sample in the case of the transmission set-up and above the sample in the case of the reflection set-up as detailed in the text.}
	\label{RBL}
\end{figure}

The THz-TDS system was used for super-resolution experiments in transmission and reflection confocal configurations (Fig.~\ref{RBL}). 
In both cases, THz radiation generated by the emitting PCA (Tx) was collected and focused to a 1~mm diameter pinhole (I) using polymethyl-pentene (TPX) lenses (L$_1$ and L$_2$), with a nominal focal length of 50~mm, to obtain a Gaussian beam. After the pinhole, an HRFZ-Si lens with a 25~mm focal length (L$_3$) collimates the THz radiation.

In transmission experiments (Fig.~\ref{RBL}(a)), another HRFZ-Si lens with 25~mm focal length (L$_4$) and numerical aperture NA=0.58, focus the radiation to the sample (S), which is a slit or a photolithographic mask (see later in the text). The samples were inserted using a sample holder orthogonal to the THz beam line axis. THz radiation diffracted by the sample was collected and focused to the receiving antenna using two TPX lenses (L$_5$ and L$_6$), with a nominal focal length of 50~mm. 

For edge experiments, the sample holder of the photolithographic mask was mounted in a bidirectional translation stage allowing movement in horizontal and vertical directions (x and y in Fig.~\ref{RBL}(a)), orthogonal to the THz beam. For the slit experiment, instead, the receiving antenna (Rx) was mounted in the
bidirectional translation stage allowing its movement in horizontal and vertical directions orthogonal to the THz beam, to measure the off-axis diffracted field.

For image acquisitions in transmission, the temporal scan range of the THz-TDS optical delay line was 40~ps, and data were pitched every 33~fs. The THz signal was averaged with an integration time of 2 s over 16 scans (each lasting 0.125 s).

In the reflection configuration (Fig.\ref{RBL}(b)), an HRFZ-Si beam-splitter (Bs) is positioned after lens L$_3$. The collimated THz radiation, once forward-transmitted, is focused onto the sample (S) using a gold-coated parabolic mirror (PM) with a nominal focal length of 50~mm and numerical aperture NA=0.45. The radiation scattered by the sample is collected by the parabolic mirror and then reflected by the beam-splitter towards another TPX lens (L$_4$) with a nominal focal length of 50~mm, which focuses the radiation onto the detection antenna (Rx).

The samples were placed in a holder that was mounted in a bidirectional translation stage for mapping, allowing movement in an xy plane orthogonal to the incident THz beam (inset of Fig.\ref{RBL}(b)). 
The temporal scan range for reflection was 50~ps, while data pitching and acquisition time remained the same as in the transmission set-up. 

To position the blade of the KE, a holder was created by 3D printing in polylactic acid (PLA): the project file was realized by the 3D CAD Tinkercad (www.tinkercad.com) and built on an Ultimaker (The Netherland) S3 3D printer. The holder includes two 45-degree angled flanges made of two layers where the blades can be introduced and tight (inset of Fig.~\ref{RBL}(b)). 
A steel shaving blade was chosen for the knife edge scan of the mask and the slit (Fig.~\ref{fig:mask-slit},
while a blade made of cardboard material 200 $\mu$m thick, coated with aluminium on one side, was employed to avoid damaging the paper samples. The blade holder was moved by a step motor, composed of one driver and one actuator/stage for each direction of movement. The control unit DC Drive T-Cube (TDC001) and the stage Z812 are both from Thorlabs (USA) and have step precision greater than 0.05~$\mu$m, much smaller than the used bidirectional translation stage step size of 10~$\mu$m. A linear translation stage driven by a micrometer with 0.01~mm graduations allows the positioning of the blade at the desired distance from the sample.

\subsection{Image acquisitions and data analysis}
For imaging acquisition, two LabView (National Instruments Corp., USA) programs were developed: the scanning module that allows the synchronism between the sample's raster scan and the acquisition of the THz signals, and the imaging module with the super-resolved image reconstruction as described in the \ref{par:KE-method} paragraph. 

The scanning module serves to scan the sample and the two blades in the x and y directions orthogonal to the THz beamline. Further, the scanning module automatically saves time domain (TD) data and
spectral data obtained by fast Fourier transform (FFT) for every blade position \cite{FP}.

The imaging module calculates the square modulus of the integral of the TD pulses, or the square modulus of the integral of the spectral signal centered at the frequency of interest and with an interval of $\pm$0.05~THz. 
In the acquisitions without the KE, these values are directly represented as pixel intensity in the final image. In the super-resolved images by using the KE, the values calculated at each blade position are subtracted from that collected at the previous position, for each pixel in the image, and the maximum difference is represented as the intensity of the corresponding image pixel.

\subsection{Samples study}
\label{par:lithografic_mask}

\begin{figure}[h]
	\centering
	\includegraphics[height=4.5 cm]{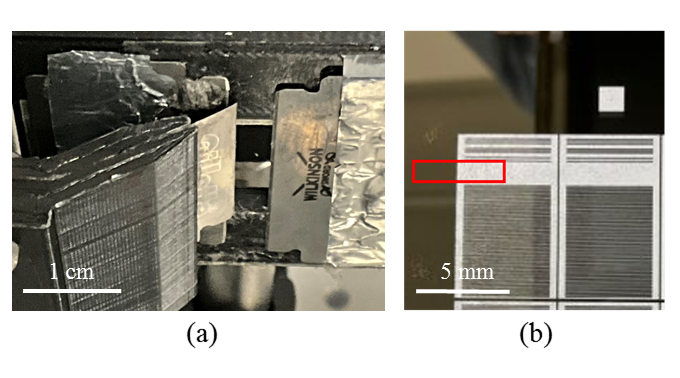}
		\caption{(a) Metal slit with 5~mm of separation and the blade of the KE placed in front of it. (b) Photography of a portion of the lithographic mask, with the red rectangle representing the 5$\times$1.5~mm$^2$ scanned area of interest in the edge experiment.}
	\label{fig:mask-slit}
\end{figure}

Super-resolution was tested on diffraction from a slit composed of two metal blades using the THz-TDS transmission setup. The blades were positioned with a separation of 5.0~mm (Fig.\ref{fig:mask-slit}(a)). 
A multimeter set to measure the electrical resistance was used to check the contact position between
the blade and the slit: the contact was verified by short-circuit conditions, and the linear translation stage driven by the micrometer was used to adjust their distance. The blade of the KE was placed at distances of 150 $\mu$m from the slit, at an angle of 45 degrees with respect to the slit plane. THz signals were detected as the KE was scanned along the slit.

Super-resolution in THz-TDS transmission mode imaging was also tested by using an optical lithographic mask (size 100×100×1.70~mm$^3$) made of a substrate of fused silica covered with a 30-nm-thick chromium layer shaped by electron-beam lithography into different-sized stripes and squares. The reflectivity of the chromium film in the THz range is $\simeq 0.99$ \cite{evan}, and therefore its transmittance is $\simeq 0.01$. The transmittance of the fused silica plate is $\simeq 0.8$ at normal incidence \cite{Naftaly2021}, hence providing strong optical image contrast.

Before the experiments, the lithography mask was examined and imaged by using an optical microscope to measure the sizes of the stripes and squares and their spacing. The acquired microscope images were calibrated by using a stage micrometer and analyzed by the ImageJ software (imagej.net/ij/) during the photo editing process.
For super-resolved imaging experiments, the lithographic mask was mounted in the THz beamline by using a sample holder realized in PLA by the Ultimaker (Nederland) S3 3D printer.

\subsection{Resolution test targets}

Test targets were realized on paper substrates to investigate THz super-resolution in the reflection configuration. Paper is made of cellulose fibres derived from plants or wood \cite{Di_Napoli_2020}, and it is known to have a low refractive index $n\sim 1.35$, and low absorption coefficient ($\le 15$ cm$^{-1}$) for frequencies below 1 THz \cite{RHcel,celpeak, Bardon_2017_part1, Paraipan_2023}. As a result, its reflectivity obtained for the case of normal incidence by using the Fresnel equations \cite{fowles2012} is $\simeq 0.02$.
On paper substrates, stripes of different materials were drawn. 

An experiment was carried out by drawing graphite stripes on a $t\sim 100~\mu$m thick paper substrate. Graphite is a semi-metal. i.e. a conductor along the plane of each carbon sheet. It is thus expected to have a high refractive index, hence providing a strong optical contrast to the paper substrate and making this kind of mockup appropriate for tests of THz super-resolved imaging \cite{Bardon_2017_part1, evan}.

To draw the graphite stripes on paper, a stencil made of PLA was realized by using the Ultimaker S3 3D printer. 
The stencil allowed the drawn graphite stripes with widths of 2~mm with a gap of 1.5~mm, which were used for the experiments.

A single mockup was prepared with carbon black 2×2~mm$^2$ squares. The objective of this mockup was to evaluate the performance of THz super-resolution in 2D imaging. This approach involved employing two blade-direction movements to obtain super-resolution in both $x$ and $y$ direction (inset of Fig.\ref{RBL}(b)).

The coherence of the THz radiation in THz-TDS causes 
multiple reflections within samples with flat and parallel surfaces
that generate echoes following the main THz pulse at a temporal distance of $\Delta t_{se}=2tn/c$, where $c$ is the speed of light. For this paper substrate, however, the first echo peak should have appeared at $\Delta t_{se}=0.9$~ps, and it was superimposed on the main peak which is $\sim 1$~ps wide\cite{FP}.

Mockup targets were also created on paper with a thickness of $t\sim 400~\mu$m using pigments commonly encountered in materials of works of art and documents. They were obtained from the company Poggi, Italy (www.poggi1825.it):  minium or red lead, whose chemical formula is Pb$_3$O$_4$ (Lead tetroxide-42008), carbon black (PBk7), cinnabar or vermilion with chemical formula $\alpha$HgS (HgS-PR106), 
and Naples Yellow with chemical formula Pb$_2$Sb$_2$O$_7$ (Cadmium sulphide PY35 + Iron Oxide PY14).

To be deposited on the paper substrate, the pigments were mixed with Arabic gum. The ratio of 1 part gum to 2 parts pigment ensured effective binding and adhesion to the paper. To achieve a smooth consistency, a few drops of distilled water were added to each mixture. In the specific case of the cinnabar, produced by the company Kremer Pigmente, Germany (kremer-pigmente.de), the pigment paste was already pre-mixed with the Arabic gum binder.
To get sharp stripes of pigments masking adhesive tape was used. Once pigments were dried, the masking tape was carefully removed, taking care not to disturb the pigmented stripes.

Mockups were created to assess the response and super-resolution imaging capabilities of the THz set-up when two different pigments are in contact with each other. In this case, the mockups consisted of two adjacent rectangles, each measuring 10×30~mm$^2$, with no separation between them. To ensure precise boundaries, the same masking adhesive tape procedure used for the striped mockups was used. The selected pigments for painting each box were cinnabar and Naples yellow, chosen because of their different transmittance in the frequency range from 0.3 to 2~THz \cite{book}. 
 
The first echo peak for the mockup targets created on paper with a thickness of $\sim 400\mu$m appeared at $\Delta t_{se}=3.6$~ps, and it was well separated from the main peak. To receive the reflection from the front surface of the pigments deposited on the paper, in data analysis the echo signal was eliminated by truncating the THz time trace after the main peak \cite{Jepsen, FP}.

The dimensional features of all these targets were analyzed by using dimension-calibrated microscope images, and the uncertainty was found to be less than 0.1~mm.

\subsection{Ancient Sample}
\label{par:parchment}
\begin{figure}[t]
	\centering
	\includegraphics[height=4cm]{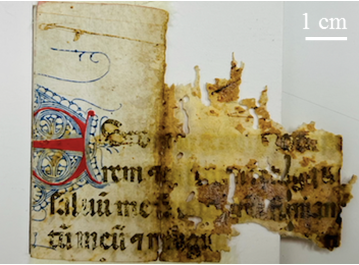}
	\caption{The medieval parchment fragment, dated back 14th century A.D.}
	\label{Parchment}
\end{figure}

The ancient sample is a medieval parchment fragment (size about 7.5x6~cm$^2$) dated back from the late 14th century A.D., kindly granted by the company Alta Formazione Italia Glocal Services Srl Impresa Sociale (Rome, Italy), and the Vatican Apostolic Library (Rome, Italy), featuring text and graphic signs (Fig. \ref{Parchment}). 

Parchment is prepared from wet unhaired animal skin soaked in a lime solution (Ca(OH)$_2$) and primarily consists of low-density collagen. To improve the smoothness and suitability of the parchment skin for writing it is mechanically smoothed and coated with organic materials. This process ensured that the writing materials did not penetrate deeply into the parchment fibre mesh \cite{cool_parchment}. 

The drop case of the text is colored with red and blue pigments and black ink, which were identified using the diagnostic technique called Fiber Optics Reflectance Spectroscopy (FORS) \cite{Aceto_2014, missori_NC_2016, Sbroscia_2020}. These substances were found to belong to the class of commonly used medieval inks and pigments based on minerals and organic salts, such as cinnabar, ultramarine, iron-gall, or carbon-based inks \cite{Aceto_2014, book, Marconi_2024}. 
The drop cap representing the letter "E" on the fragment measures approximately 3~cm in height, and it is larger than the standard text, reaching a height of up to about 0.5~cm. 

The fragment appears heavily deteriorated, exhibiting some dirty incrustations. Additionally, a layer of glue applied to preserve the material appears to be deteriorating. All of these features of the fragment were considered in the setup of the THz imaging experiment to make it as non-invasive as feasible.
Accordingly, a sample holder was devised from cardboard, matching the medieval parchment fragment's dimensions and featuring a 2~cm$\times$4~cm aperture for the study area of interest centered on the drop cap. This shows features ranging from 0.5~mm to 2~cm in size and includes all types of colors, allowing for a thorough evaluation of the capabilities of the super-resolution THz imaging set-up.

\section{Results and discussion}

\subsection{Standard resolution of the THz beam lines}
\label{par:standard-resolution}

In this section, the expected resolution without KE is calculated for the transmission and reflection THz beam lines used for the experiments. The THz radiation emitted by the PCA has an approximately Gaussian profile \cite{Jepsen_1995, ke1, Tofani_2019_IEEE}, and in the case of the transmission beamline (Fig.~\ref{RBL}(a)) is collimated by the L$_1$ lens in a beam with a diameter ($D$) of approximately 20~mm. This is focused by the L$_2$ lens into a spot, which according to the formula for the spot size of Gaussian beams $2W_0$ is \cite{bahaa_2019}:

\begin{equation}
2W_0 = \frac{4  f  c}{\pi D \nu}
\label{W0}
\end{equation}

where $f$ is the focal length of L$_2$ lens.
For the frequency $\nu=0.3$~THz this formula gives $2W_0=3.18$~mm. 
This spot is focused on the plane of an iris with a diameter $d_i=2.2$~mm, which produces a diffraction pattern of the transmitted radiation. This is collected by the L$_3$ lens. The first minimum of the Fraunhofer diffraction pattern of a circular aperture is at the radial distance $\rho$ from the optical axis of:

\begin{equation}
\rho= f \sin (\frac{\theta_0}{2})=f \sin (\frac{1.22 \lambda}{2 d_i})
\end{equation}

where $\theta_0$ is the angular diameter of the first minimum of the diffraction pattern and $f$ is the focal length of L$_3$ lens. Using the beamline parameters, we obtain the value $\theta_0$=33.68$^{\circ}$ and $\rho$=7.24~mm. Therefore the central maximum of the diffraction pattern created by the iris has a diameter of 14.48~mm. In the Gaussian beam approximation, and by using the Eq.~\ref{W0}, the 0.3~THz radiation is focused by the L$_4$ lens into a 2.28~mm diameter spot on the slit or the lithographic mask.
When using the reflection THz beamline (Fig.~\ref{RBL}(b)) the PM focuses the 0.3~THz radiation into a 1.82~mm diameter spot on the mockup targets and the 0.6~THz radiation into a 0.91~mm diameter spot. 
According to the Rayleigh criterion \cite{fowles2012}, these values are the expected resolution that can be obtained without the KE.

\subsection{Slit experiment}

In the THz-TDS transmission setup, a 5~mm aperture slit (Fig. \ref{fig:mask-slit}(a)) was used to explore super-resolution through images of its diffraction pattern and diffraction profile without and with the KE. At a subwavelength distance between the blade and the slit, the evanescent-to-propagating wave conversion is expected to happen \cite{evan}. Therefore, the KE was placed at approximately 150~$\mu$m from the slit (Fig. \ref{fig:mask-slit}(a)), and the THz signal at 0.3~THz ($\lambda=1$~mm) was considered in the analysis.

The slit causes the diffraction of the THz radiation, which is collected and collimated by the L$_5$ lens placed at its focal distance from the slit. The collimated THz beam is focused by the L$_6$ lens together with the hemispherical Si lens to the dipole of the receiver PCA (Rx of Fig. \ref{RBL}(a)). The L$_6$ lens is placed approximately 20~mm from the hemispherical Si lens of the PCA. 
To acquire the diffraction pattern, the Rx device was horizontally translated by $\pm$10.5~mm relative to the optical axis of the transmission setup at two symmetrical vertical positions, with x-step=0.5~mm and y-step=1~mm.  
When used, the blade of the KE was scanned parallel to the edge of the slit, for a length of 5~mm with 0.2~mm spatial steps.

\begin{figure}[h!]
	
	\centering
	\begin{subfigure}
		\centering
		\includegraphics[width=0.5\columnwidth]{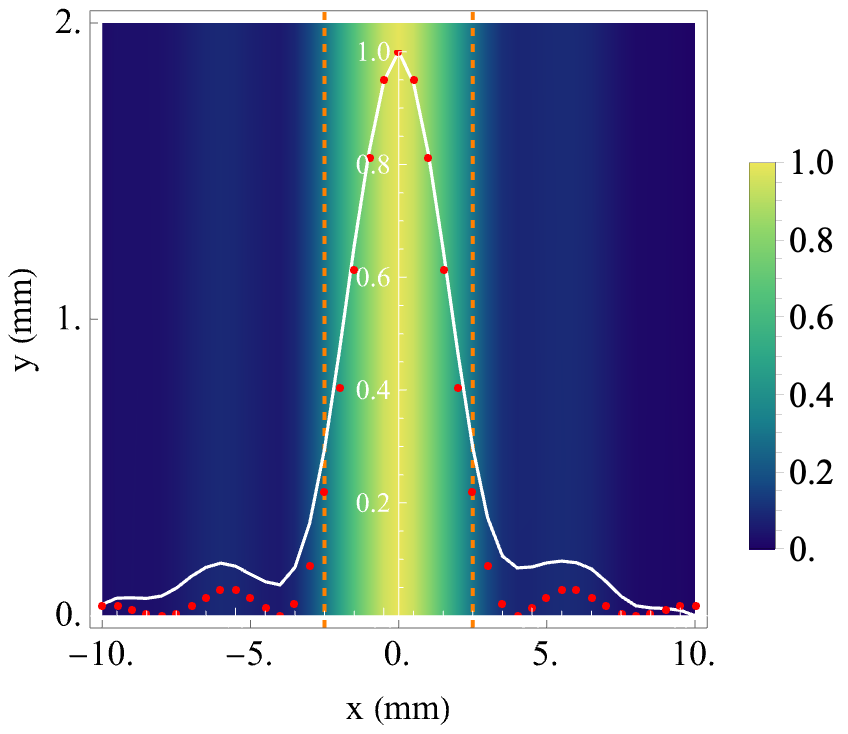}\\{(a)}
	\end{subfigure}

	\vspace{0.5cm}
	\begin{subfigure}
		\centering
		\includegraphics[width=0.5\columnwidth]{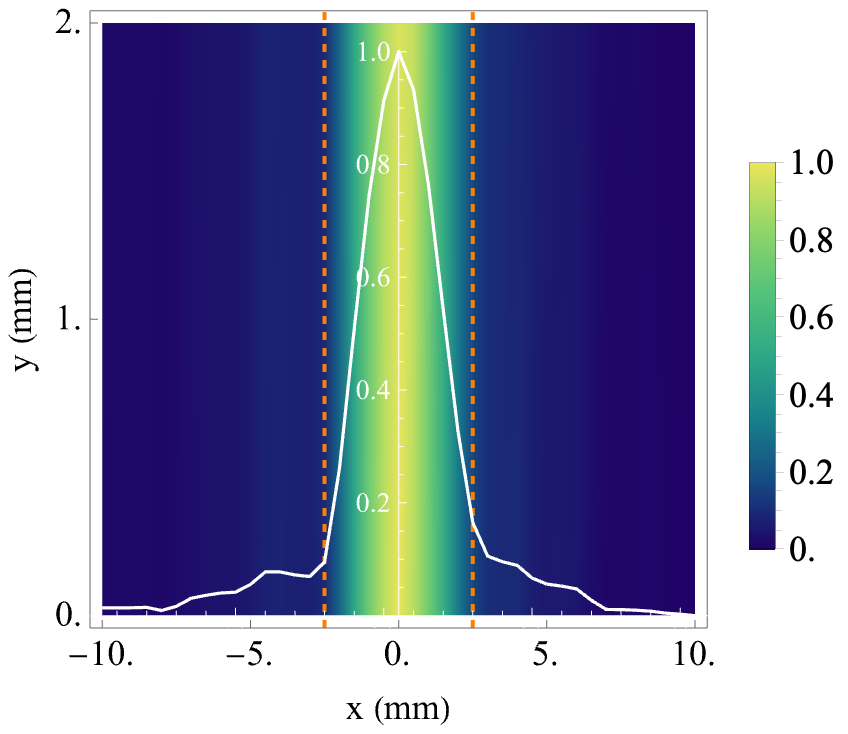}\\{(b)}
	\end{subfigure}

	\caption{THz-TDS transmission images (21~mm$\times$2~mm) at 0.3~THz  of the 5~mm aperture slit obtained without (a) and with (b) the KE. In the panels are also shown the 0.3~THz intensity profiles along x obtained by averaging the two horizontal scans of each image (white curves) and the opening of the slit by the two dotted vertical orange lines. In panel (a) the red dots are analytical simulations of the diffraction pattern.} 
	\label{0.3THz}
\end{figure}

The THz diffraction images without and with the KE are shown in Fig.~\ref{0.3THz}(a) and (b), respectively. The corresponding intensity profiles along x, obtained by averaging the pixel intensity of the two horizontal scans of each image and normalizing the result to 1, are shown superimposed to the images as white curves. 

%\subsection{Simulations}
Analytical simulations were conducted using Mathematica Wolfram software to calculate the diffraction pattern created by the 5~mm slit in the THz beamline.
This was obtained by calculating the far-field diffraction pattern as:
\begin{equation}
	I=\frac{e^{-ikd}e^{-ik\frac{x_s^2}{2d}}}{d}\int_{- \infty}^{\infty} \alpha(x)e^{ik\frac{x_sx}{d}}dx
\end{equation}
where $k$ is the wave vector of the radiation and $x_s$ is the x coordinate on the screen.
This expression stems directly from the Huygens-Fresnel principle as the sum, in the paraxial approximation, of the secondary spherical waves generated when a plane wave is incident on an aperture $\alpha(x)$  placed perpendicular to the propagation direction $z$, at distance $d$ from the screen. For a slit with aperture $a_s$, $\alpha(x)=\theta(a_s/2-x)\theta(a_s/2+x)$.

In the absence of the KE, a Fraunhofer diffraction pattern was observed in the far-field region (Fig.~\ref{0.3THz}(a)). The obtained image featured a central peak and two side lobes, which nicely agree with analytical simulations of the diffraction without the KE shown as red dots in Fig.~\ref{0.3THz}(a). By using the KE, the Fraunhofer approximation is no longer valid and the Fresnel diffraction regime appears, where the central peak is narrower and the side lobes tend to disappear, as shown in Fig.~\ref{0.3THz}(b) \cite{bahaa_2019}. In addition, the width at half maximum of the diffraction profile without the KE is 3.85~mm, while with the KE it reduces to 3.13~mm, a 1.23-fold decrease.

\subsection{Edge experiment}
Having observed Fresnel diffraction in the far field, we applied the KE method to enhance resolution when imaging realistic objects. We first consider an opaque edge of the lithographic mask.
With this aim in mind, we captured an image measuring 5.0~mm~$\times$~1.5~mm (with x-step=0.2~mm and y-step=0.5~mm) depicting a sharp boundary between the chromium layer and the fused silica substrate of the lithographic mask (Fig.~\ref{fig:mask-slit}(a)) \cite{evan}.
The sharp boundary was scanned with and without the KE. When used, the blade of the KE was scanned parallel to the edge at a distance of 200~$\mu$m from the mask, for a length of 5~mm with 0.2~mm spatial steps.

\begin{figure*}[tp]
	
	\centering
	\begin{subfigure}
		\centering
		\includegraphics[width=0.35\linewidth]{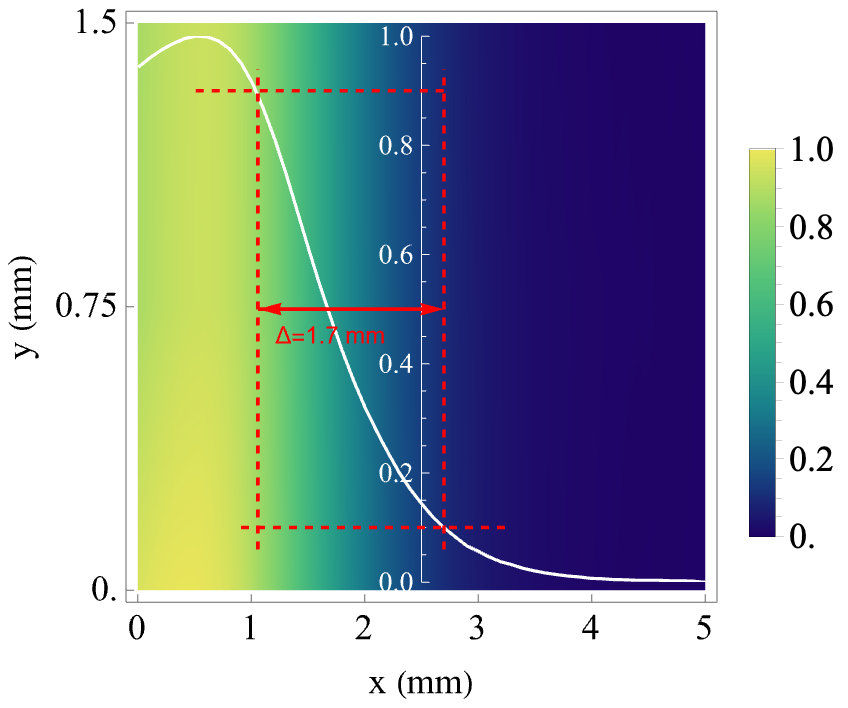}
        \includegraphics[width=0.35\linewidth]{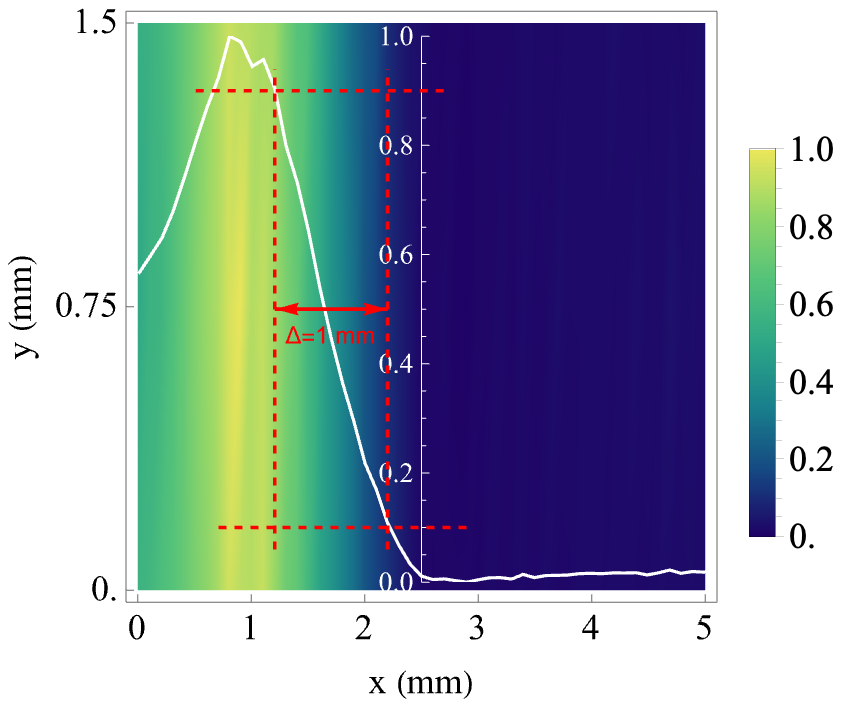}\\{(a)} \hspace{7cm} {(b)}
	\end{subfigure}

	\vspace{0.5cm}
	\begin{subfigure}
		\centering
		\includegraphics[width=0.35\linewidth]{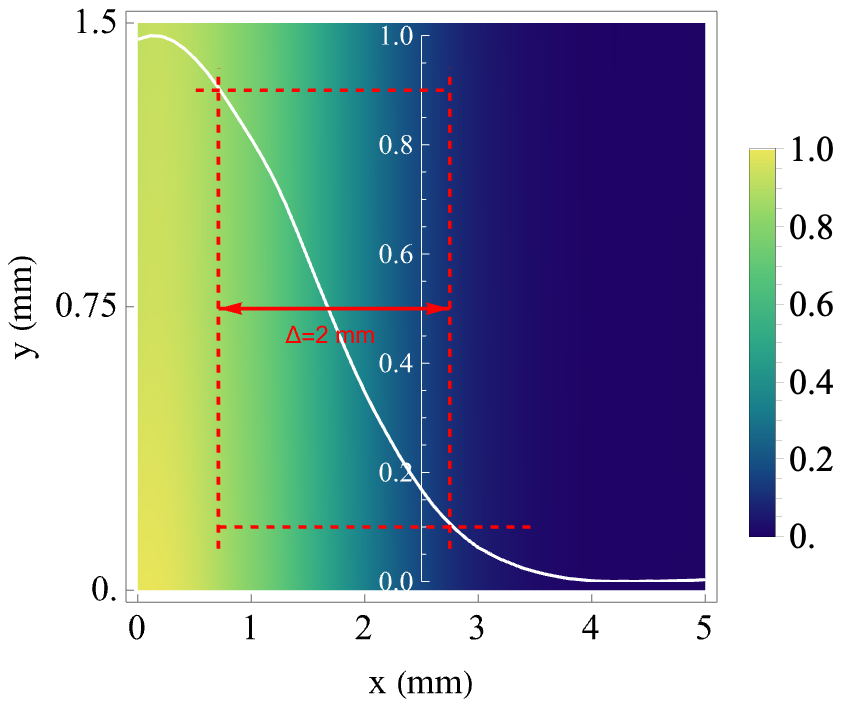}
        \includegraphics[width=0.35\linewidth]{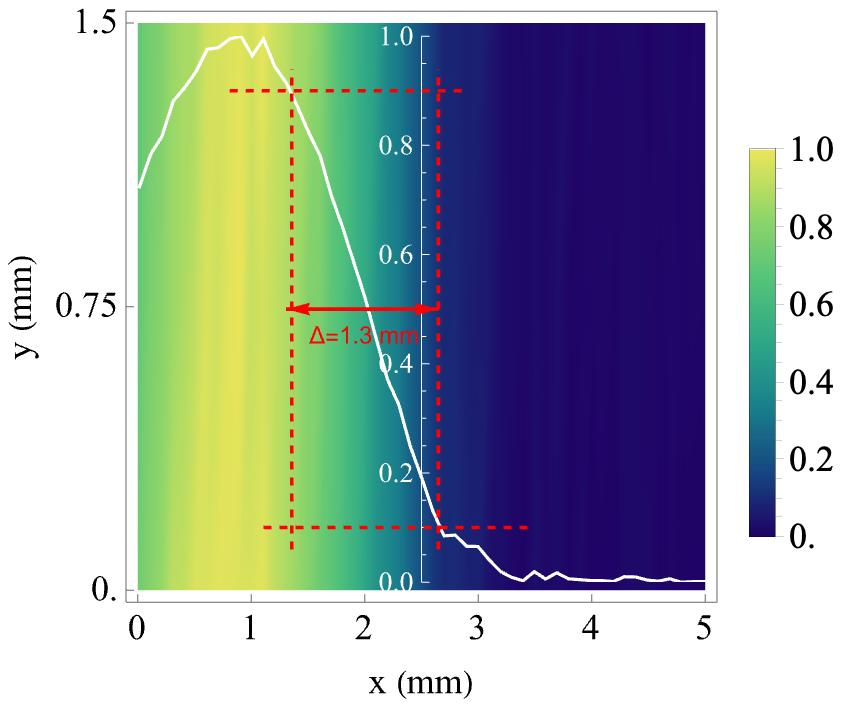}\\{(c)} \hspace{7cm} {(d)}
	\end{subfigure}

	\caption{THz images of a chromium border area (5.5~mm$\times$1.5~mm) of the lithographic mask (Fig.~\ref{fig:mask-slit}): a) TD image; b) TD image with the KE; c) image at 0.3~THz; d) image at 0.3~THz with the KE. In the panels are also shown the THz intensity profiles along x obtained by averaging the horizontal scans of each image (white curves) and the calculation of the slope.} 
	\label{Mask}
\end{figure*}

Fig.s~\ref{Mask}(a)-(d) shows the obtained THz images: those of panels (a) and (b) are obtained from the time-domain (TD) pulses, thus containing all the propagating frequencies of the broadband THz radiation; panels (c) and (d), instead, show the images obtained at 0.3~THz. As evident from the images, the inclusion of the KE, panels (b) and (d), made the images sharper in both cases. In addition, it is possible to appreciate the appearance of the oscillating Fresnel diffraction pattern at a boundary, although only the first maximum in the diffraction profile in super-resolved images can be observed.

The normalized THz signal intensity profiles along the x-axis were obtained by using the pixel intensities averaged on the three different scans at different y coordinates, with and without the KE (white lines in Fig.~\ref{Mask}(a)-(d)).

The imaging resolution was assessed by calculating the slope of the normalized THz signal intensity profiles in variation from 0.1 to 0.9. The profile for TD without the KE show a slope of 0.47~mm$^{-1}$, corresponding to an Abbe diffraction limit of 2.13~mm. With KE, a slope of 0.8~mm$^{-1}$, corresponding to a subwavelength resolution of approximately 1.25~mm, is obtained, with an enhancement of spatial resolution of 1.7~times.

Profiles along the x-axis obtained at 0.3~THz (Fig.~\ref{Mask}(c)) also demonstrated resolution improvement.  Without the KE, a slope of 0.4~mm$^{-1}$, corresponding to an Abbe diffraction limit of 2.5~mm, is found, in good agreement with calculations of Section~\ref{par:standard-resolution}. A slope of 0.62~mm$^{-1}$, corresponding to a resolution of about 1.63~mm, is obtained, with an enhancement of spatial resolution of 1.53~times.

These results are in agreement with a previous study using a continuous-wave single-frequency THz source \cite{evan}. It was reported a progressive transition from near-field imaging to diffraction-limited imaging when the structured illumination plane by the KE is lifted from the object plane up to a distance on the order of one wavelength.
Our results demonstrate the effectiveness of imaging below the Rayleigh resolution limit with KE by using a THz-TDS setup.

\subsection{Imaging of graphite features on paper}
\label{par:graphite_on_paper}

\begin{figure}[h]
	
	\centering
	\begin{subfigure}
		\centering
		\includegraphics[width=0.5\columnwidth]{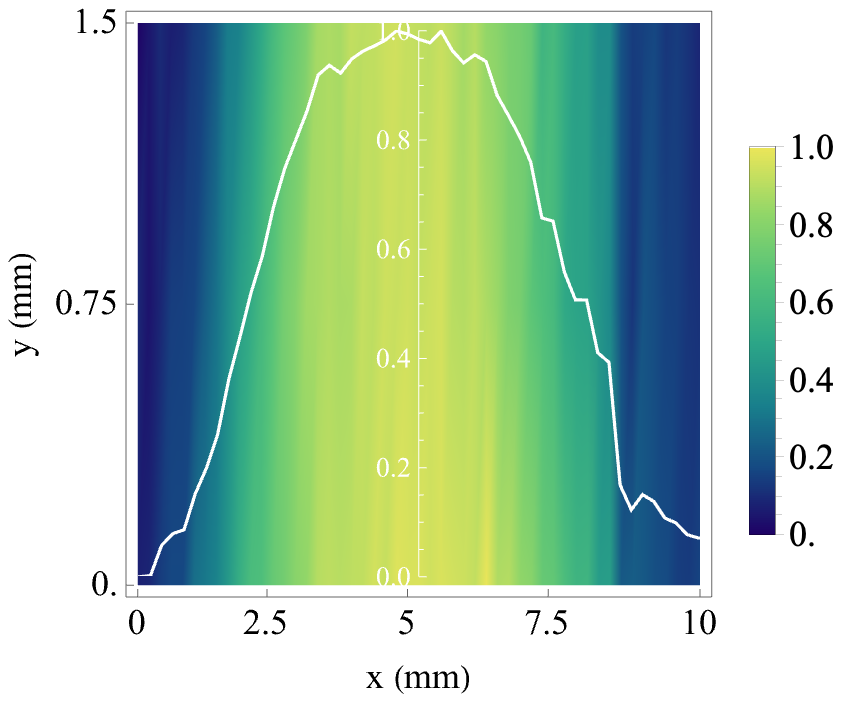}\\{(a)}
	\end{subfigure}

	\vspace{0.5cm}
	\begin{subfigure}
		\centering
		\includegraphics[width=0.5\columnwidth]{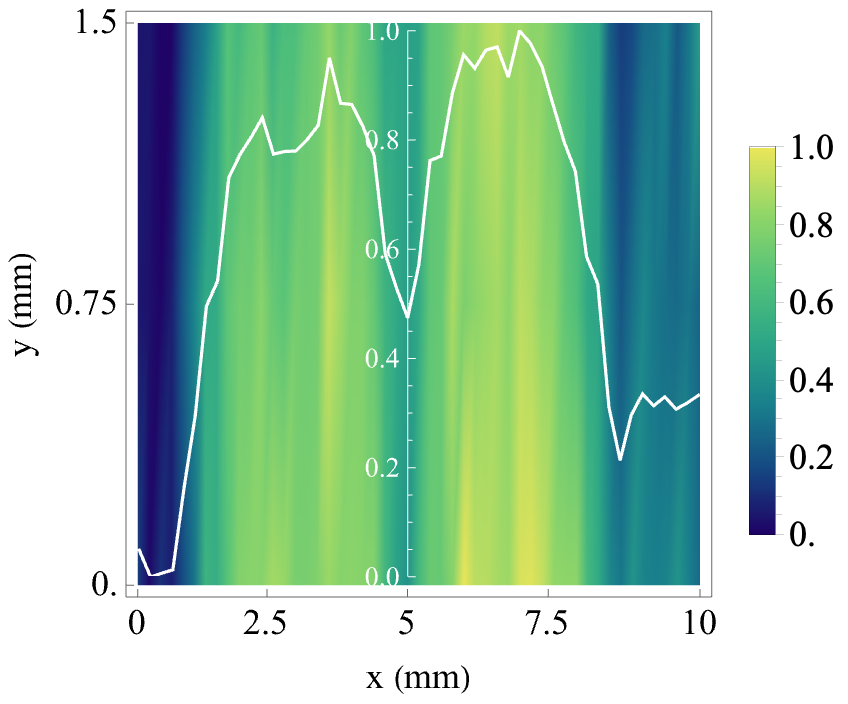}\\{(b)}
	\end{subfigure}

	\caption{Images obtained at 0.3~THz by the reflection setup (Fig.~\ref{RBL}(b)) of graphite stripes on paper, 2~mm wide with a gap of 1.5~mm. (a) Image obtained without KE; (b) image obtained with KE. In the panels are also shown the THz intensity profiles along x obtained by averaging the horizontal scans of each image (white curves).} 
	\label{G03}
\end{figure}

Fig.s~\ref{G03}(a) and (b) show images 10.5~mm$\times$1.5~mm wide (x-step=0.2~mm, y-step=0.5~mm) obtained at 0.3~THz using the reflection setup (Fig.~\ref{RBL}(b)) demonstrating the impact of the KE scan on visualization of the graphite stripes, 2~mm wide with a gap of 1.5~mm, on paper. When used, the blade made of cardboard material coated with aluminium was scanned in contact with the paper surface, parallel to the graphite stripes, for a length of 5~mm with 0.2~mm spatial steps.
The aluminum-coated side of the cardboard was positioned facing upwards, effectively creating a 200~$\mu$m gap between the blade (the aluminum) and the sample surface.

As calculated in Section \ref{par:standard-resolution}, the resolution achievable by the reflection setup without the KE is 1.82 mm. This value is corroborated by the image in Fig.~\ref{G03}(a), acquired without the KE, revealing a single broad feature with the two graphite stripes being indistinguishable.
Instead, the violation of the Rayleigh criterion is evident in Fig.~\ref{G03}(b), where the two stripes are discernible, thus confirming the effectiveness of KE super-resolved imaging in reflection.

Profiles along the x-axis (Fig.~\ref{G03}) quantitatively show the resolution improvement. Without the KE technique, only a broad-intensity band is observed. In contrast, when using the KE, two distinguishable features can be observed, with a minimum between them falling to approximately 50\% of the maximum.

These results validate the super-resolution approach using the THz-TDS reflection setup, paving the way for exploring the improvement of resolution when imaging other materials, such as pigments and inks in documental cultural heritage.

\subsection{Imaging of pigments on paper}

\subsubsection{Carbon black stripes mockup}

The THz reflectivity of carbon black pigment provides a strong optical contrast to the paper substrate, similar to graphite \cite{Bardon_2013, Bardon_2017_part1}.
The analysis focused on 1~mm-wide carbon black stripes, separated by 2~mm. Images of 8.0~mm$\times$0.4~mm (x-step=0.1~mm, y-step=0.2~mm) were obtained using the reflection setup. 
When used, the blade was scanned with the same parameters used for graphite stripes on paper (section \ref{par:graphite_on_paper}).

\begin{figure}[h]
	
	\centering
	\begin{subfigure}
		\centering
		\includegraphics[width=0.5\columnwidth]{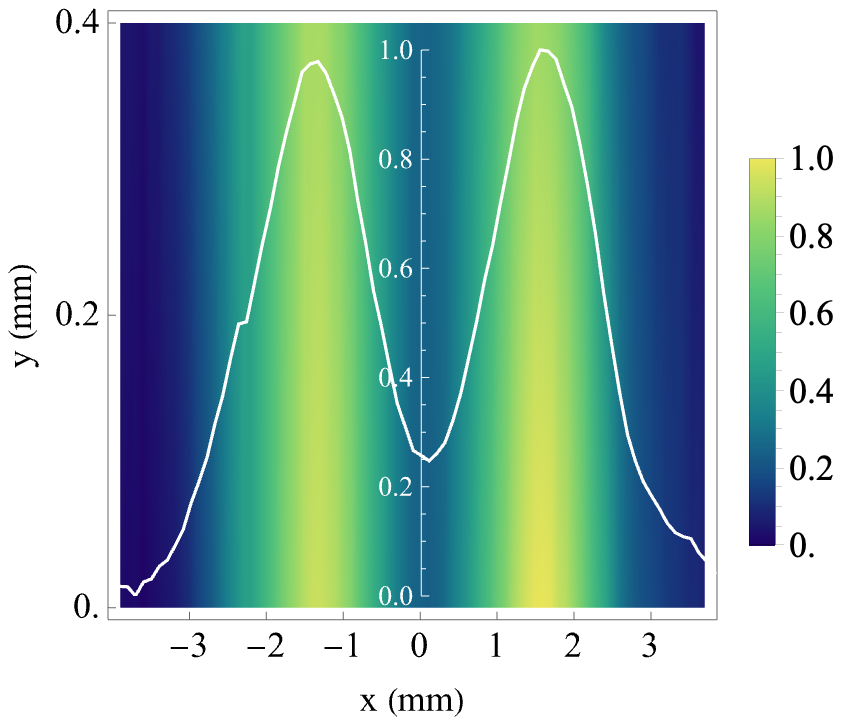}\\{(a)}
	\end{subfigure}

	\vspace{0.5cm}
	\begin{subfigure}
		\centering
		\includegraphics[width=0.5\columnwidth]{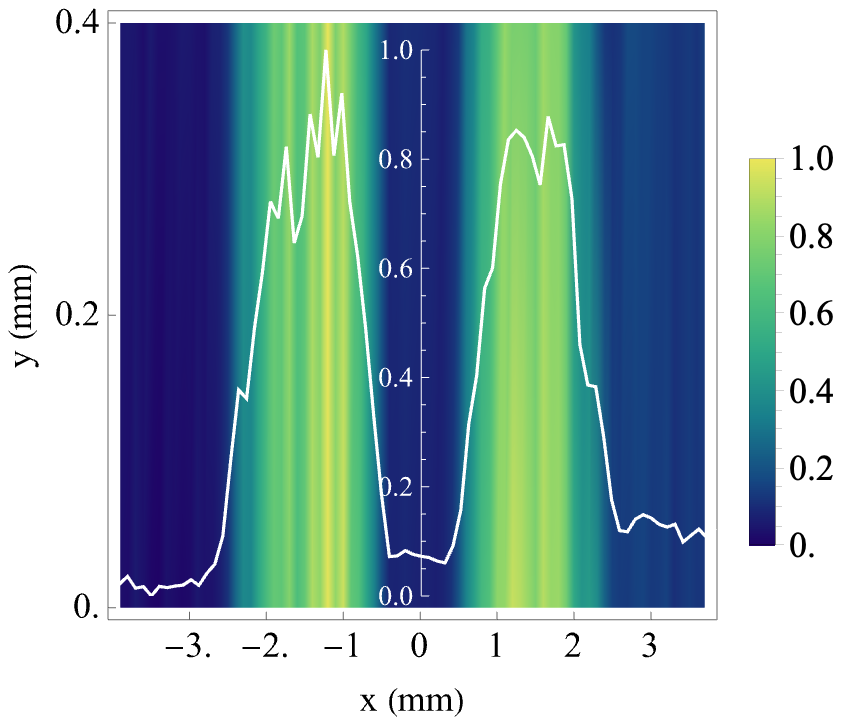}\\{(b)}
	\end{subfigure}

	\caption{Images at 0.3~THz of the carbon black mockup with stripes  1~mm wide with a gap of 2~mm: a) without KE, b) with KE. In the panels are also shown the THz intensity profiles along x obtained by averaging the horizontal scans of each image (white curves).} 
	\label{CB2I}
\end{figure}

The image without the KE (Fig.~\ref{CB2I}(a)) consists of two distinguishable stripes, as expected due to the resolution of 1.82~mm of the reflection setup without the KE obtained in the Section~\ref{par:standard-resolution}. In the case of super-resolution (Fig.~\ref{CB2I}(b)), the image of the two strips appears sharper and it is possible to appreciate irregularities in the intensity of the reflected signal, most likely due to the irregularities in the application of the pigment on the paper substrate.

Profiles along the x-axis quantitatively show the resolution improvement. Without the KE technique, two smooth intensity peaks are observed, with a minimum between them falling to approximately 25\% of the maximum. When using the KE the two peaks appear sharper and with much more roughness.
The concentration of pigment influences the contrast of the image, with a higher concentration on the paper surface resulting in a greater amplitude of the reflected waveform \cite{Bardon_2017_part1}. Additionally, the topographic features of graphic signs, which can arise from surface irregularities of the paper substrate, may also affect the contrast of THz images \cite{bardon_2017_part2}.
The two features appear much better separated, with the minimum between them falling to approximately 7\% of the maximum.

\subsubsection{Minium stripes mockup}

Minium (Pb$_3$O$_4$) can be considered as a semiconductor and lead to well-contrasted images on the paper substrate due to its larger refractive index (close to 2) with respect to that of paper (about 1.35) \cite{RHcel,celpeak,Bardon_2017_part1}.

\begin{figure}[h]
	
	\centering
	\begin{subfigure}
		\centering
		\includegraphics[width=0.5\columnwidth]{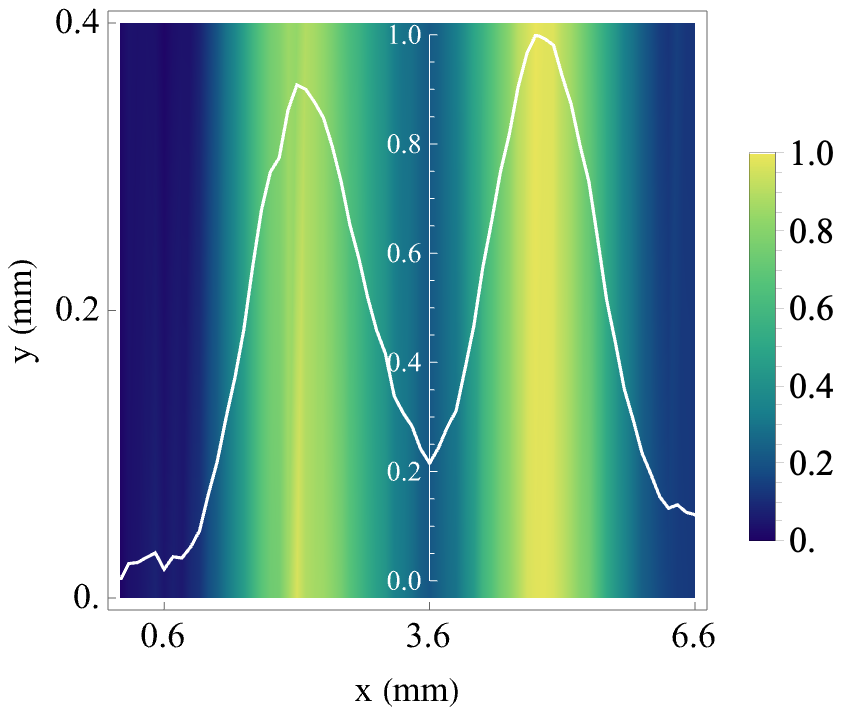}\\{(a)}
	\end{subfigure}

	\vspace{0.5cm}
	\begin{subfigure}
		\centering
		\includegraphics[width=0.5\columnwidth]{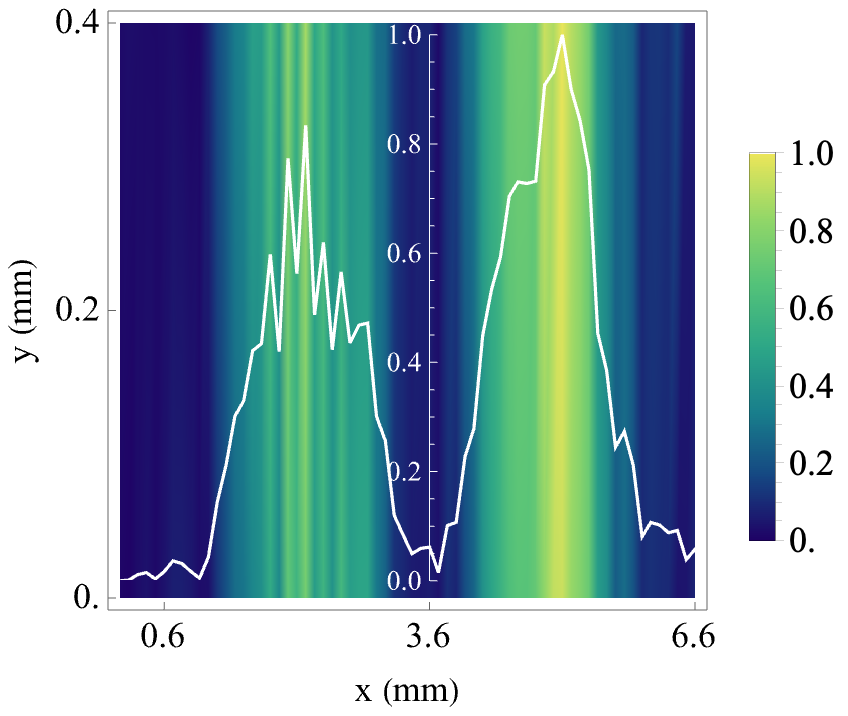}\\{(b)}
	\end{subfigure}

	\caption{THz images at 0.3~THz of two minium 1~mm~wide stripes separated by 2~mm: a) image without KE; b) image with KE. In the panels are also shown the THz intensity profiles along x obtained by averaging the horizontal scans of each image (white curves).} 
	\label{CnI}
\end{figure}

Figure~\ref{CnI} presents images of 4.6~mm$\times$0.4~mm (x-step=0.1~mm, y-step=0.2~mm) and profiles of two minium 1~mm~wide stripes separated by 2~mm obtained without and with KE. When used, the blade was scanned with the same parameters used for graphite stripes on paper (section III-D).

In a similar way to the images of the carbon black pigment, the image without the KE (Fig.~\ref{CnI}(a)) consists of two distinguishable stripes, while using the super-resolution (Fig.~\ref{CnI}(b)), the image of the two strips appears sharper and it is possible to appreciate irregularities in the intensity of the reflected signal, most likely due to the concentration inhomogeneities of the pigment \cite{Bardon_2017_part1}.
Profiles along the x-axis quantitatively show the resolution improvement passing from two smooth peaks, with a minimum between them falling to approximately 20\% of the maximum, to two sharper and irregular peaks when using the KE. In this last case, the two features appear much better separated, with the minimum between them falling to approximately 5\% of the maximum.

\subsubsection{Cinnabar and Naples yellow stripes in contact}

The experiment aimed to assess THz imaging's resolution capabilities of the boundary between two pigments in contact: the selection of cinnabar and Naples yellow pigments for this experiment was driven by their substantial differences in reflectivity. Cinnabar has a lower transmittance and lower reflectance than Naples yellow in the THz frequency range used in this work (http://thzdb.org). 

\begin{figure}[h]
	
	\centering
	\begin{subfigure}
		\centering
		\includegraphics[width=0.4\columnwidth]{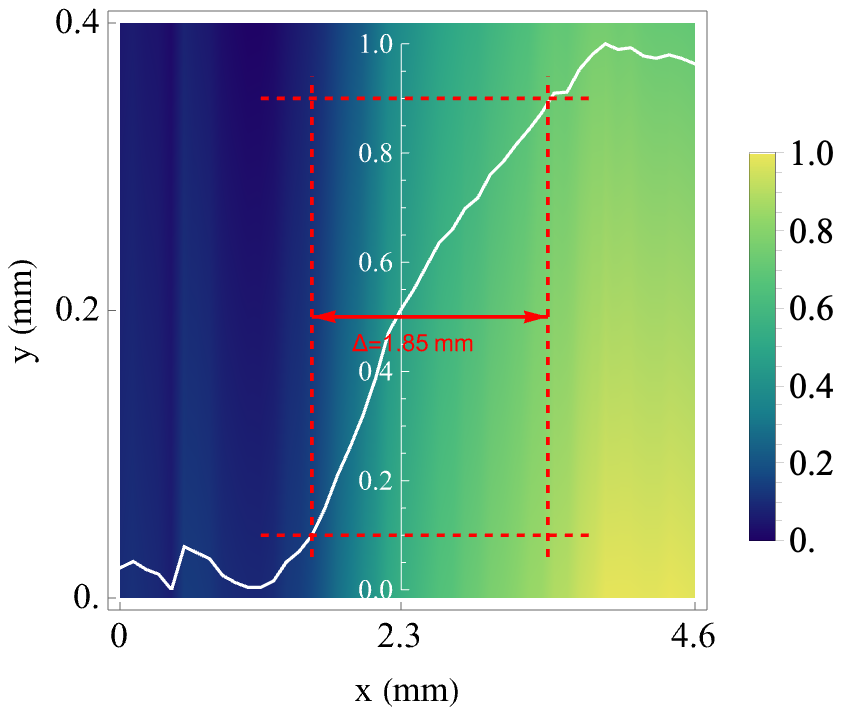}\\{(a)}
	\end{subfigure}

	\vspace{0.5cm}
	\begin{subfigure}
		\centering
		\includegraphics[width=0.4\columnwidth]{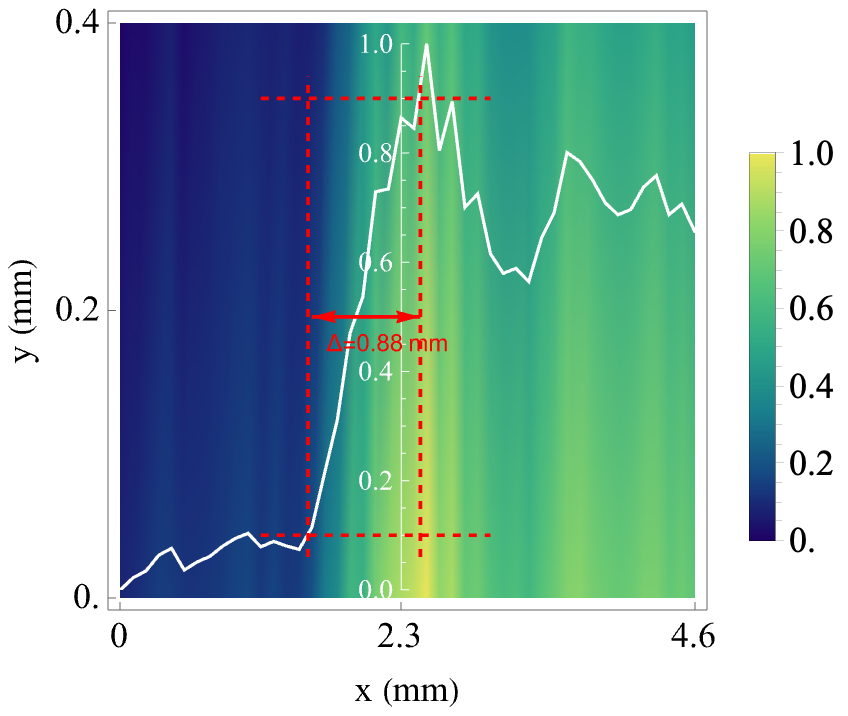}\\{(b)}
	\end{subfigure}

	\caption{THz images at 0.3~THz of cinnabar and Naples yellow pigments in contact: a) image without KE, b) image with KE. In the panels are also shown the THz intensity profiles along x obtained by averaging the horizontal scans of each image (white curves) and the calculation of the slope.} 
	\label{CnNYI}
\end{figure}

Images and intensity profiles at 0.3~THz, with and without the KE, of a transition region between the two pigments placed in the centre of the mockup, are shown in Fig.~\ref{CnNYI}. The size of the image is 4.6~mm$~\times~$0.4~mm (x-step=0.1~mm, y-step=0.2~mm), the transition between the two pigments occurs along the x coordinate. %and their frequencies are 0.3 and 0.6~THz. 
When used, the blade was scanned parallel to the boundary with the same parameters used for graphite stripes on paper (section III-D).

The imaging resolution was assessed by calculating the slope of the normalized THz signal intensity profiles in variation from 0.1 to 0.9. The profiles without the KE show a slope of 0.43~mm$^{-1}$, corresponding to an Abbe diffraction limit of 2.31~mm. With KE, a slope of 0.91~mm$^{-1}$, corresponding to a resolution of approximately 1.1~mm, is obtained. A resolution improvement of a factor 2.1 is observed and a Fresnel intensity profile associated with the transition between the two pigments is evident for the super-resolved image in Fig.~\ref{CnNYI}(b).

\subsubsection{2D imaging of a carbon black square mockup}

\begin{figure*}[tp]
	
	\centering
	\begin{subfigure}
		\centering
		\includegraphics[width=0.35\linewidth]{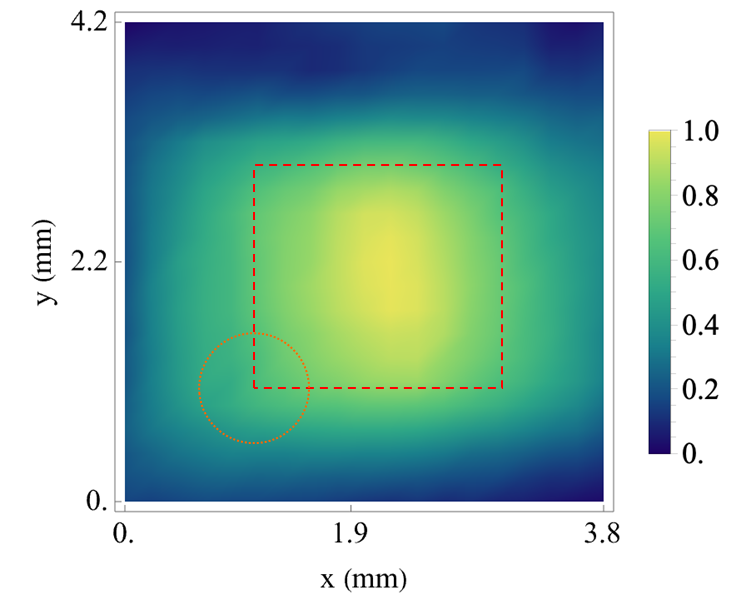}
        \includegraphics[width=0.35\linewidth]{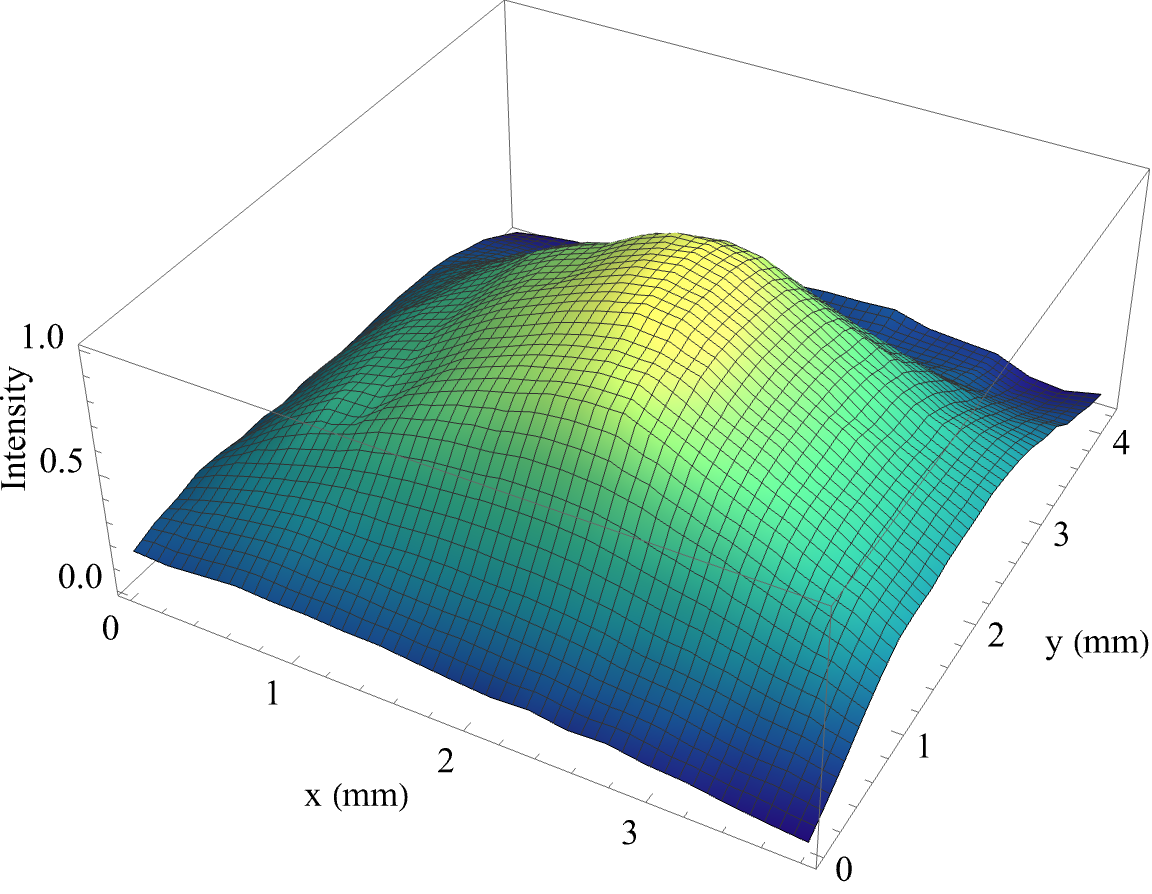}\\{(a)} \hspace{7cm} {(b)}
	\end{subfigure}

	\vspace{0.5cm}
	\begin{subfigure}
		\centering
		\includegraphics[width=0.35\linewidth]{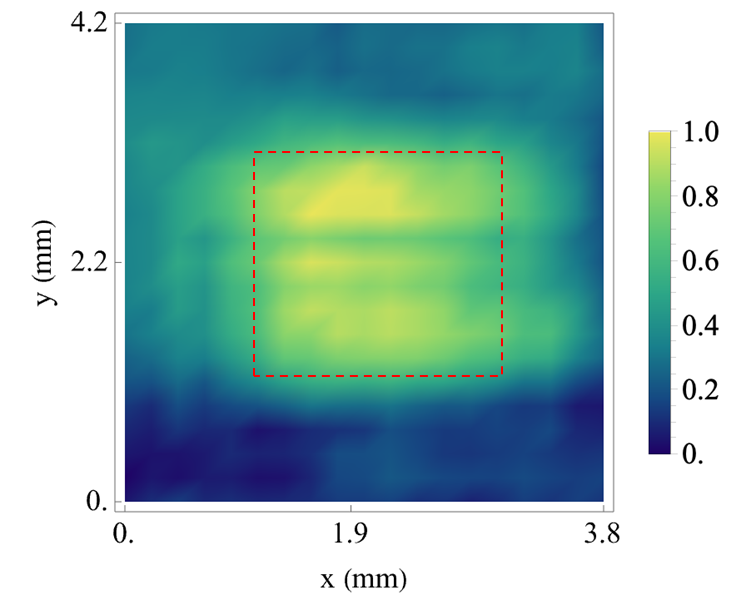}
        \includegraphics[width=0.35\linewidth]{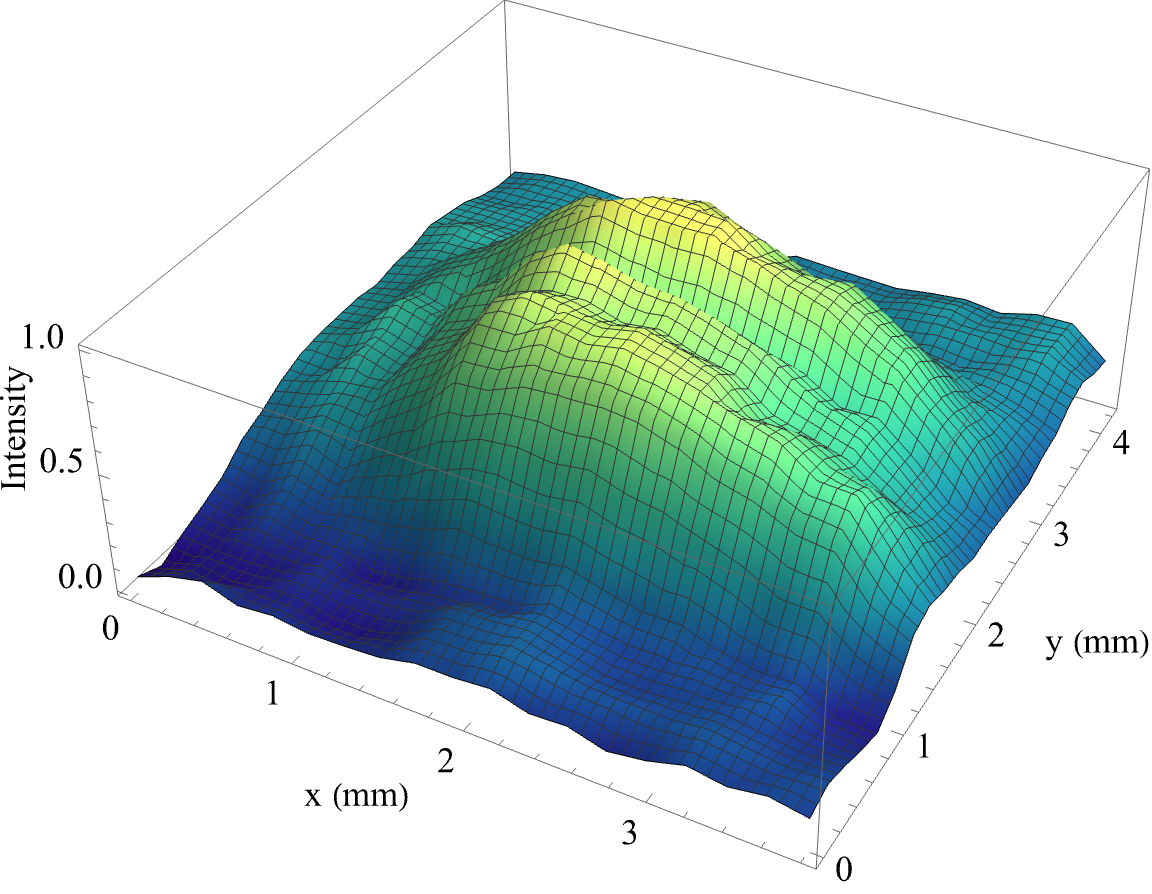}\\{(c)} \hspace{7cm} {(d)}
	\end{subfigure}

	\caption{THz images at 0.6~THz of the 2~mm$\times$2~mm carbon black square on paper: a) image without KE and b) 3D graph of the intensity; c) image with KE and d) 3D graph of the intensity. The squares dashed in red in panels (a) and (c) represent the carbon black square. The orange circle in panel (a) represents the THz beam spot size.} 
	\label{CBsqI}
\end{figure*}

In the cases illustrated above, the mockups were essentially one-dimensional structures with spatial discontinuities along the x-axis.
In this section, we describe the application of KE super-resolution technique for a two-dimensional structure with spatial discontinuities along the x and y-axis. The imaging was performed on a 2~mm$\times$2~mm carbon black square on paper.  
The area under investigation spans 3.8~mm$\times$4.2~mm, with spatial steps of 0.2~mm in both the x and y coordinates.
To enhance spatial resolution along both axes, we employed the complete KE setup depicted in the inset of Fig.~\ref{RBL}, utilizing blade movements along the x and y directions. When used, the blade made of cardboard material coated with aluminium was scanned in contact with the paper surface, parallel to the sides of the carbon black square, for a length of 3~mm in both directions and with 0.3~mm spatial steps.

Fig.~\ref{CBsqI} shows THz images of the carbon black square at 0.6~THz, with and without the KE.
Greater amplitudes of the reflected THz waveform correspond to a higher concentration of carbon black pigment \cite{Bardon_2017_part1}, since topographic irregularities are negligible in the 2$\times$2~mm area of the graphic sign \cite{Bardon_2017_part1}.

In the absence of the KE, the THz image (Fig.~\ref{CBsqI}(a)) and the 3D profile of the intensity (Fig.~\ref{CBsqI}(b)) appear to be made up of a single diffuse spot while the super-resolved ones (Fig.s~\ref{CBsqI}(c) and (d)) reveal a sharper image and details associated with the different concentrations of the carbon black pigment in the form of two distinct areas along the y-axis. These patterns serve as markers, providing insights into the structural features and inherent imperfections of the sample.

This underscores the unique capability of THz imaging to reveal nuanced details in graphic signs, presenting a distinct perspective compared to traditional optical microscopy.

\subsection{THz super-resolution imaging of the medieval manuscript}
This section describes the application of the KE super-resolution to a real ancient document, such as the medieval parchment fragment described in section \ref{par:parchment}. 
Parchment, like paper, can be regarded as a dielectric material, with a higher refractive index of approximately 1.7 compared to paper. The differences in refractive index values between the cinnabar used for the graphic signs and the surrounding support are consequently minimized when parchment is considered. Therefore, the differences in the amplitude of the THz pulse reflected, with or without the pigment, are lower for parchment \cite{Bardon_2017_part1}. 

Therefore, to increase the signal-to-noise ratio and have better contrast between pigment and substrate, the images were analyzed using TD pulses and, consequently, considering all the frequencies contained in the pulses, rather than a specific frequency.

We employed the complete KE setup depicted in the inset of Fig.~\ref{RBL}, utilizing blade movements along the x and y directions.
The imaging without KE was performed on a 10~mm$\times$6~mm area. When the KE was used, a 5~mm$\times$4~mm image was acquired. A spatial step of 0.2~mm in both the x and y coordinates was used for both images. When using the KE, the blade was scanned for a length of 3~mm in both directions and with 0.3~mm spatial steps.
 
\begin{figure}[h!]
	\centering
	\includegraphics[height=6.3cm]{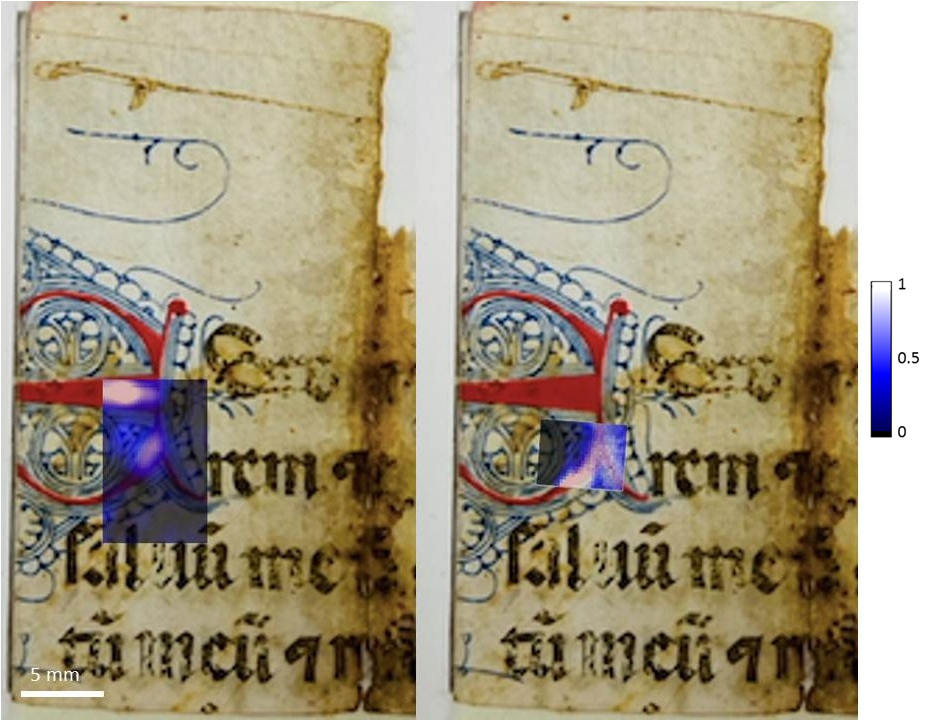}
	\caption{Superimposed to the visible images of the medieval parchment are shown the THz images (inserted rectangles) of areas at the initial drop cap of the text. THz image without the KE (left), with the KE (center), and the THz intensity scale normalized to 1 (right).}
	\label{fig:parchment}
\end{figure}

The images were targeted at the initial drop cap of the text, which represents the letter "E", whose main red graphic strokes are of cinnabar. The THz images obtained are shown superimposed to those visible in Fig.~\ref{fig:parchment}.

The image without the KE shows two large, shapeless features corresponding to the broad graphic lines drawn with the red cinnabar pigment in the drop cap.
The super-resolved image with KE represents a significant improvement in spatial resolution and fidelity where the brightest regions indicated areas with higher concentrations of cinnabar pigment \cite{Bardon_2017_part1}. 
The super-resolved image achieved the remarkable feat of capturing features of cinnabar that were less than 1~mm in size.
The ultramarine graphic signs are not discernible even in the super-resolved image due to the transparency of this pigment in the THz frequencies range \cite{book}.

\section{Conclusion}

This work focuses on improving the inherent spatial resolution limitations in THz-TDS imaging caused by far-field diffraction. The proposed super-resolution approach relies on structured illumination techniques operating at a distance shorter than the wavelength from the object plane. This is achieved through a mechanical KE scan with one or two blades, within the transmission or reflection configuration of the THz-TDS beamline.

The validity of the super-resolution KE method was preliminarily demonstrated by obtaining super-resolution in a classical diffraction experiment with a slit in a transmission geometry. A transition from Fraunhofer to Fresnel diffraction patterns was observed with the decreasing width at half maximum of the diffraction pattern by a factor of 1.23. We then obtained an improvement in spatial resolution in imaging of a boundary between a transmissive and an opaque zone in a lithographic mask. This demonstrated the super-resolution capability of the KE setup, with an enhancement of spatial resolution of 1.7 times at the boundary.

Next, we considered a reflection configuration of the THz-TDS beamline with a single-axis KE profiling. Resolution targets made of two parallel graphite stripes, each 2 mm wide and separated by a distance of 1.5 mm, were imaged, demonstrating the possibility of resolving the two stripes within a violation of the Rayleigh limit. Super-resolved spatial features were also obtained by imaging mockup targets made using the pigments carbon black, cinnabar, minium, and Naples yellow on a paper substrate.

The KE method was also applied to 2D objects using two blades moving in orthogonal directions to image a carbon black square measuring 2 mm × 2 mm. The results further demonstrated spatial super-resolution, providing insights into the inhomogeneity in the pigment concentration within the square. The 2D super-resolution method was also applied to the imaging of a medieval manuscript. The KE method significantly enhanced spatial resolution, enabling the capture of graphic features smaller than 1 mm.

Our work has successfully demonstrated the ability to visualize structures beyond the Rayleigh resolution limit by using a mechanical KE scan. This achievement significantly expands the applicative scope of THz-TDS imaging while enhancing its practicality and efficacy.

\section*{Acknowledgments}
We are grateful to Silvia Sotgiu, Véronique Cachia, and Giuliano Locatelli de Maestri of the National Central Library of Rome, Italy, and Federica Delia of Recto Verso, Rome, Italy, for help in mockup preparations.
DAV also thanks funding provided by the Erasmus Mundus scholarship.

\bibliographystyle{IEEEtran}
\bibliography{bibliografia}

\begin{thebibliography}{10}
\providecommand{\url}[1]{#1}
\csname url@samestyle\endcsname
\providecommand{\newblock}{\relax}
\providecommand{\bibinfo}[2]{#2}
\providecommand{\BIBentrySTDinterwordspacing}{\spaceskip=0pt\relax}
\providecommand{\BIBentryALTinterwordstretchfactor}{4}
\providecommand{\BIBentryALTinterwordspacing}{\spaceskip=\fontdimen2\font plus
\BIBentryALTinterwordstretchfactor\fontdimen3\font minus \fontdimen4\font\relax}
\providecommand{\BIBforeignlanguage}[2]{{%
\expandafter\ifx\csname l@#1\endcsname\relax
\typeout{** WARNING: IEEEtran.bst: No hyphenation pattern has been}%
\typeout{** loaded for the language `#1'. Using the pattern for}%
\typeout{** the default language instead.}%
\else
\language=\csname l@#1\endcsname
\fi
#2}}
\providecommand{\BIBdecl}{\relax}
\BIBdecl

\bibitem{Abbe}
E.~Abbe, ``Beiträge zur theorie des mikroskops und der mikroskopischen wahrnehmung,'' \emph{Archiv für Mikroskopische Anatomie}, vol.~9, no.~1, pp. 413--468, dec 1873.

\bibitem{Juang_1988}
C.-B. Juang, L.~Finzi, and C.~J. Bustamante, ``Design and application of a computer-controlled confocal scanning differential polarization microscope,'' \emph{Review of Scientific Instruments}, vol.~59, no.~11, pp. 2399--2408, Nov. 1988.

\bibitem{book}
K.~Fukunaga, \emph{THz Technology Applied to Cultural Heritage in Practice}.\hskip 1em plus 0.5em minus 0.4em\relax Springer, 2016.

\bibitem{Yang_2016}
X.~Yang, X.~Zhao, K.~Yang, Y.~Liu, Y.~Liu, W.~Fu, and Y.~Luo, ``Biomedical applications of terahertz spectroscopy and imaging,'' \emph{Trends in Biotechnology}, vol.~34, no.~10, pp. 810--824, Oct. 2016.

\bibitem{Wang_2017}
K.~Wang, D.-W. Sun, and H.~Pu, ``Emerging non-destructive terahertz spectroscopic imaging technique: Principle and applications in the agri-food industry,'' \emph{Trends in Food Science \& Technology}, vol.~67, pp. 93--105, Sep. 2017.

\bibitem{Wang_2021}
Q.~Wang, L.~Xie, and Y.~Ying, ``Overview of imaging methods based on terahertz time-domain spectroscopy,'' \emph{Applied Spectroscopy Reviews}, vol.~57, no.~3, pp. 249--264, Jan. 2021.

\bibitem{Koch_2023}
M.~Koch, D.~M. Mittleman, J.~Ornik, and E.~Castro-Camus, ``Terahertz time-domain spectroscopy,'' \emph{Nature Reviews Methods Primers}, vol.~3, no.~1, Jun. 2023.

\bibitem{Totero_Gongora_2020}
J.~S. Totero~Gongora, L.~Olivieri, L.~Peters, J.~Tunesi, V.~Cecconi, A.~Cutrona, R.~Tucker, V.~Kumar, A.~Pasquazi, and M.~Peccianti, ``Route to intelligent imaging reconstruction via terahertz nonlinear ghost imaging,'' \emph{Micromachines}, vol.~11, no.~5, p. 521, May 2020.

\bibitem{Yuan}
F.~Yuan, S.~Zhou, T.~Cheng, and Y.~Liu, ``A low- and high-resolution terahertz image pair construction method with gradient fusion for learning-based super-resolution,'' \emph{{IEEE} Access}, vol.~10, pp. 132\,506--132\,514, 2022.

\bibitem{Lei}
T.~Lei, B.~Tobin, Z.~Liu, S.-Y. Yang, and D.-W. Sun, ``A terahertz time-domain super-resolution imaging method using a local-pixel graph neural network for biological products,'' \emph{Analytica Chimica Acta}, vol. 1181, p. 338898, oct 2021.

\bibitem{Ljubenovi_2023}
M.~Ljubenović, A.~Artesani, S.~Bonetti, and A.~Traviglia, ``Super-resolution of thz time-domain images based on low-rank representation,'' in \emph{2023 Sixth International Workshop on Mobile Terahertz Systems (IWMTS)}.\hskip 1em plus 0.5em minus 0.4em\relax IEEE, Jul. 2023.

\bibitem{Yu}
T.~Yu, X.~Zuo, W.~Liu, and C.~Gong, ``0.1thz super-resolution imaging based on 3d printed confocal waveguides,'' \emph{Optics Communications}, vol. 459, p. 124896, mar 2020.

\bibitem{SIM}
X.~Chen, S.~Zhong, Y.~Hou, R.~Cao, W.~Wang, D.~Li, Q.~Dai, D.~Kim, and P.~Xi, ``Superresolution structured illumination microscopy reconstruction algorithms: a review,'' \emph{Light: Science and Applications}, vol.~12, no.~1, jul 2023.

\bibitem{Strohl:16}
F.~Ströhl and C.~F. Kaminski, ``Frontiers in structured illumination microscopy,'' \emph{Optica}, vol.~3, no.~6, p. 667, Jun. 2016.

\bibitem{Gusta}
M.~G.~L. Gustafsson, ``Surpassing the lateral resolution limit by a factor of two using structured illumination microscopy. {SHORT} {COMMUNICATION},'' \emph{Journal of Microscopy}, vol. 198, no.~2, pp. 82--87, may 2000.

\bibitem{Firester1977}
\BIBentryALTinterwordspacing
A.~H. Firester, M.~E. Heller, and P.~Sheng, ``Knife-edge scanning measurements of subwavelength focused light beams,'' \emph{Appl. Opt.}, vol.~16, no.~7, pp. 1971--1974, Jul 1977. [Online]. Available: \url{https://opg.optica.org/ao/abstract.cfm?URI=ao-16-7-1971}
\BIBentrySTDinterwordspacing

\bibitem{sub_2015}
S.~H. Phing, A.~Mazhorova, M.~Shalaby, M.~Peccianti, M.~Clerici, A.~Pasquazi, Y.~Ozturk, J.~Ali, and R.~Morandotti, ``Sub-wavelength terahertz beam profiling of a {THz} source via an all-optical knife-edge technique,'' \emph{Scientific Reports}, vol.~5, no.~1, feb 2015.

\bibitem{metamaterials}
J.~F. O{\textquotesingle}Hara, R.~Singh, I.~Brener, E.~Smirnova, J.~Han, A.~J. Taylor, and W.~Zhang, ``Thin-film sensing with planar terahertz metamaterials: sensitivity and limitations,'' \emph{Optics Express}, vol.~16, no.~3, p. 1786, 2008.

\bibitem{confocal}
C.~Ciano, M.~Flammini, V.~Giliberti, P.~Calvani, E.~DelRe, F.~Talarico, M.~Torre, M.~Missori, and M.~Ortolani, ``Confocal imaging at 0.3 {THz} with depth resolution of a painted wood artwork for the identification of buried thin metal foils,'' \emph{{IEEE} Transactions on Terahertz Science and Technology}, vol.~8, no.~4, pp. 390--396, jul 2018.

\bibitem{Burford_2017}
N.~M. Burford and M.~O. El-Shenawee, ``Review of terahertz photoconductive antenna technology,'' \emph{Optical Engineering}, vol.~56, no.~1, p. 010901, Jan. 2017.

\bibitem{FP}
R.~Fastampa, L.~Pilozzi, and M.~Missori, ``Cancellation of fabry-perot interference effects in terahertz time-domain spectroscopy of optically thin samples,'' \emph{Physical Review A}, vol.~95, no.~6, p. 063831, jun 2017.

\bibitem{evan}
M.~Flammini, E.~Pontecorvo, V.~Giliberti, C.~Rizza, A.~Ciattoni, M.~Ortolani, and E.~DelRe, ``Evanescent-wave filtering in images using remote terahertz structured illumination,'' \emph{Physical Review Applied}, vol.~8, no.~5, p. 054019, nov 2017.

\bibitem{Naftaly2021}
M.~Naftaly and A.~Gregory, ``Terahertz and microwave optical properties of single-crystal quartz and vitreous silica and the behavior of the boson peak,'' \emph{Applied Sciences}, vol.~11, p. 6733, 07 2021.

\bibitem{Di_Napoli_2020}
B.~Di~Napoli, S.~Franco, L.~Severini, M.~Tumiati, E.~Buratti, M.~Titubante, V.~Nigro, N.~Gnan, L.~Micheli, B.~Ruzicka, C.~Mazzuca, R.~Angelini, M.~Missori, and E.~Zaccarelli, ``Gellan gum microgels as effective agents for a rapid cleaning of paper,'' \emph{ACS Applied Polymer Materials}, vol.~2, no.~7, pp. 2791--2801, May 2020.

\bibitem{RHcel}
M.~Missori, D.~Pawcenis, J.~Bagniuk, A.~M. Conte, C.~Violante, M.~Maggio, M.~Peccianti, O.~Pulci, and J.~{\L}ojewska, ``Quantitative diagnostics of ancient paper using {THz} time-domain spectroscopy,'' \emph{Microchemical Journal}, vol. 142, pp. 54--61, nov 2018.

\bibitem{celpeak}
M.~Peccianti, R.~Fastampa, A.~M. Conte, O.~Pulci, C.~Violante, J.~{\L}ojewska, M.~Clerici, R.~Morandotti, and M.~Missori, ``Terahertz absorption by cellulose: Application to ancient paper artifacts,'' \emph{Physical Review Applied}, vol.~7, no.~6, p. 064019, jun 2017.

\bibitem{Bardon_2017_part1}
T.~Bardon, R.~K. May, J.~B. Jackson, G.~Beentjes, G.~de~Bruin, P.~F. Taday, and M.~Strlič, ``Contrast in terahertz images of archival documents—part i: Influence of the optical parameters from the ink and support,'' \emph{Journal of Infrared, Millimeter, and Terahertz Waves}, vol.~38, no.~4, pp. 443--466, Jan. 2017.

\bibitem{Paraipan_2023}
A.~A. Paraipan, N.~Luchetti, A.~Mosca~Conte, O.~Pulci, and M.~Missori, ``Low-frequency vibrations of saccharides using terahertz time-domain spectroscopy and ab-initio simulations,'' \emph{Applied Sciences}, vol.~13, no.~17, p. 9719, Aug. 2023.

\bibitem{fowles2012}
G.~R. Fowles, \emph{Introduction to modern optics}.\hskip 1em plus 0.5em minus 0.4em\relax Dover Publications, Inc., New York, 2012, chapt. 2.

\bibitem{Jepsen}
P.~U. Jepsen, D.~G. Cooke, and M.~Koch, ``Terahertz spectroscopy and imaging - modern techniques and applications,'' \emph{Laser Photonics Rev.}, vol.~5, no.~1, pp. 124--166, 2011.

\bibitem{cool_parchment}
\BIBentryALTinterwordspacing
W.~Henry and et~al., ``Parchment treatments,'' in \emph{Paper Conservation Catalog}.\hskip 1em plus 0.5em minus 0.4em\relax Washington D.C.: American Institute for Conservation Book and Paper Group., 1988, ch.~18. [Online]. Available: \url{http://cool.conservation-us.org/coolaic/sg/bpg/pcc/18_parchment.pdf}
\BIBentrySTDinterwordspacing

\bibitem{Aceto_2014}
M.~Aceto, A.~Agostino, G.~Fenoglio, A.~Idone, M.~Gulmini, M.~Picollo, P.~Ricciardi, and J.~K. Delaney, ``Characterisation of colourants on illuminated manuscripts by portable fibre optic uv-visible-nir reflectance spectrophotometry,'' \emph{Analytical Methods}, vol.~6, no.~5, p. 1488, 2014.

\bibitem{missori_NC_2016}
M.~Missori, ``Optical spectroscopy of ancient paper and textiles,'' \emph{Il Nuovo Cimento C}, vol.~39, pp. 1--10, 2016.

\bibitem{Sbroscia_2020}
M.~Sbroscia, M.~Cestelli-Guidi, F.~Colao, S.~Falzone, C.~Gioia, P.~Gioia, C.~Marconi, D.~Mirabile~Gattia, E.~Loreti, M.~Marinelli, M.~Missori, F.~Persia, L.~Pronti, M.~Romani, A.~Sodo, G.~Verona-Rinati, M.~Ricci, and R.~Fantoni, ``Multi-analytical non-destructive investigation of pictorial apparatuses of “villa della piscina” in rome,'' \emph{Microchemical Journal}, vol. 153, p. 104450, Mar. 2020.

\bibitem{Marconi_2024}
C.~Marconi, A.~Mosca~Conte, O.~Pulci, and M.~Missori, \emph{Multispectral Imaging and Optical Spectroscopy of Two Letters of St. Francis de Sales}.\hskip 1em plus 0.5em minus 0.4em\relax Springer Nature Switzerland, 2024, pp. 117--128.

\bibitem{Jepsen_1995}
P.~U. Jepsen and S.~R. Keiding, ``Radiation patterns from lens-coupled terahertz antennas,'' \emph{Optics Letters}, vol.~20, no.~8, p. 807, Apr. 1995.

\bibitem{ke1}
S.~Tofani, D.~C. Zografopoulos, M.~Missori, R.~Fastampa, and R.~Beccherelli, ``Terahertz focusing properties of polymeric zone plates characterized by a modified knife-edge technique,'' \emph{Journal of the Optical Society of America B}, vol.~36, no.~5, p. D88, mar 2019.

\bibitem{Tofani_2019_IEEE}
T.~Silvia, Z.~D. C., M.~Mauro, F.~Renato, and B.~Romeo, ``High-resolution binary zone plate in double-sided configuration for terahertz radiation focusing,'' \emph{IEEE Photonics Technology Letters}, vol.~31, no.~2, pp. 117--120, Jan. 2019.

\bibitem{bahaa_2019}
B.~E.~A. Saleh and M.~C. Teich, \emph{Fundamentals of photonics}.\hskip 1em plus 0.5em minus 0.4em\relax John Wiley \& Sons, 2019, chap. 2.

\bibitem{Bardon_2013}
T.~Bardon, R.~K. May, P.~F. Taday, and M.~Strlič, ``Systematic study of terahertz time-domain spectra of historically informed black inks,'' \emph{The Analyst}, vol. 138, no.~17, p. 4859, 2013.

\bibitem{bardon_2017_part2}
T.~Bardon, R.~K. May, P.~F. Taday, and M.~Strlic, ``Contrast in terahertz images of archival documents—part ii: Influence of topographic features,'' \emph{Journal of Infrared, Millimeter, and Terahertz Waves}, vol.~38, no.~4, pp. 467--482, Jan. 2017.

\end{thebibliography}

\newpage

\vspace{11pt}

\vspace{11pt}

\vfill

\end{document}